\documentclass{article}

\usepackage{microtype}
\usepackage{graphicx}
\usepackage{subcaption}
\usepackage{booktabs} 

\usepackage{hyperref}

\usepackage[preprint]{icml2026}

\usepackage{amsmath}
\usepackage{amssymb}
\usepackage{mathtools}
\usepackage{amsthm}
\usepackage{enumitem}

\usepackage[capitalize,noabbrev]{cleveref}

\theoremstyle{plain}

\theoremstyle{definition}

\theoremstyle{remark}

\usepackage[textsize=tiny]{todonotes}

\icmltitlerunning{Scaling Open Discrete Audio Foundation Models with Interleaved Semantic, Acoustic, and Text Tokens}

\begin{document}

\twocolumn[
  \icmltitle{Scaling Open Discrete Audio Foundation Models with \\ Interleaved Semantic, Acoustic, and Text Tokens}



  \icmlsetsymbol{equal}{*}

  \begin{icmlauthorlist}
    \icmlauthor{Potsawee Manakul}{a,b}
    \icmlauthor{Woody Haosheng Gan}{d}
    \icmlauthor{Martijn Bartelds}{a}
    \icmlauthor{Guangzhi Sun}{e}
    \icmlauthor{William Held}{a,c}
    \icmlauthor{Diyi Yang}{a}
  \end{icmlauthorlist}

  \icmlaffiliation{a}{Stanford University}
  \icmlaffiliation{b}{SCB 10X}
  \icmlaffiliation{c}{OpenAthena}
  \icmlaffiliation{d}{University of Southern California}
  \icmlaffiliation{e}{University of Cambridge}

  \icmlcorrespondingauthor{Potsawee Manakul}{potsawee@stanford.edu}
  \icmlcorrespondingauthor{William Held}{held@stanford.edu}
  \icmlcorrespondingauthor{Diyi Yang}{diyiy@stanford.edu}

  \icmlkeywords{Audio Foundation Models, Scaling Laws, Speech Language Models, Next-Token Prediction, Discrete Audio}

  \vskip 0.3in
]


\printAffiliationsAndNotice{}  

\begin{abstract}
Current audio language models are predominantly text-first, either extending pre-trained text LLM backbones or relying on semantic-only audio tokens, limiting general audio modeling. 
This paper presents a systematic empirical study of native audio foundation models that apply next-token prediction to audio at scale, jointly modeling semantic content, acoustic details, and text to support both general audio generation and cross-modal capabilities.
We provide comprehensive empirical insights for building such models:
(1) We systematically investigate design choices---data sources, text mixture ratios, and token composition---establishing a validated training recipe.
(2) We conduct the first scaling law study for discrete audio models via IsoFLOP analysis on 64 models spanning $3{\times}10^{18}$ to $3{\times}10^{20}$ FLOPs, finding that optimal data grows 1.6$\times$ faster than optimal model size.
(3) We apply these lessons to train SODA (Scaling Open Discrete Audio), a suite of models 
from 135M to 4B parameters on 500B tokens, comparing against our scaling predictions and 
existing models.
SODA serves as a flexible backbone for diverse audio/text tasks---we demonstrate this by fine-tuning for voice-preserving speech-to-speech translation, using the same unified architecture.



\end{abstract}

\section{Introduction}
\label{sec:intro}

Building foundation models that can understand and generate audio is a key challenge in multimodal AI. Current approaches have distinct limitations. \textit{LLM-centric} architectures, such as SALMONN \cite{salmonn} or Qwen3-Omni \cite{qwenomni}, add audio modules to a pre-trained text LLM; while effective for instruction-following, they have a ``semantic bottleneck'' that limits general audio-to-audio modeling. \textit{Semantic-only} speech language models, such as TWIST \cite{twist} or SpiritLM \cite{spiritlm}, are trained speech-first but discard acoustic details, limiting high-fidelity understanding and generation. \textit{Native audio} models like Moshi \cite{moshi} or Llama-Mimi \cite{llamamimi} model acoustic tokens directly but focus on specific tasks without text integration. Meanwhile, next-token prediction has enabled unified models for text and vision-language \cite{chameleon}, yet analogous approaches that jointly model audio understanding and generation in a single backbone remain limited.

To bridge this gap, this paper presents a systematic empirical study of \textit{native} audio foundation models that jointly model semantic, acoustic, and text tokens within a unified next-token prediction framework---establishing the first training recipes and scaling laws analogous to scaling study in LLMs~\cite{kaplan2020}. This design enables a range of tasks within a single model: audio continuation, semantic/acoustic understanding, cross-modal capabilities (e.g., text-to-speech and speech-to-text), and text generation. We adopt utterance-level interleaving of tokens derived from neural codecs, avoiding word-level alignment errors and enabling the use of large datasets with available transcripts. A challenge in training such audio models is the lack of established pretraining understanding: while the Chinchilla study for text LLMs \cite{chinchilla} established that model size $N$ and training tokens $D$ should scale equally ($N^*, D^* \propto C^{0.5}$), it is unclear if this holds for audio, where information density per token can be far lower.
We address key questions for pre-training discrete audio models:
\vspace{-0.2cm}
\begin{itemize}[leftmargin=*]
    \setlength\itemsep{-0.3em}
    \item \textbf{What training data and token design should we use?} (\S\ref{sec:design_choice}): We systematically compare speech corpora, text mixture ratios, and token compositions (semantic-only vs.\ semantic+acoustic vs.\ semantic+acoustic+text), establishing a validated training recipe.
    \item \textbf{How should we allocate compute, and is validation loss a reliable metric?} (\S\ref{sec:scaling}): We show that validation loss is predictive of downstream performance, then derive scaling laws from 64 IsoFLOP models ($3 \times 10^{18}$ to $3 \times 10^{20}$ FLOPs), finding $D^* \propto C^{0.579}$ and $N^* \propto C^{0.367}$.
    \item \textbf{Does scaling up work?} (\S\ref{sec:largescale}): We train {SODA} ({S}caling {O}pen {D}iscrete {A}udio), a suite of models from 135M to 4B parameters on 500B tokens (up to $1.3 \times 10^{22}$ FLOPs), and validate them against our scaling predictions. We compare cold-start (from scratch) versus warm-start (from text LLMs) training at scale, finding that cold-start is superior and provides higher training stability. We further validate SODA as a flexible backbone by formulating voice-preserving speech-to-speech translation simply as a next-token prediction task and fine-tuning SODA. 
\end{itemize}
\vspace{-0.2cm}

SODA achieves competitive performance across audio and cross-modal benchmarks, with fine-tuning for S2ST demonstrating its flexibility. We release checkpoints, discrete audio data, experiment log, and code to facilitate future research.
\section{Related Work}
\label{sec:related}

\subsection{Audio \& Speech Foundation Models}

\textbf{LLM-Centric Architectures.} Models such as SALMONN \cite{salmonn}, Llama-Omni \cite{llamaomni}, and Qwen3-Omni \cite{qwenomni} warm-start from a pre-trained text LLM and add audio capability via separate encoder/decoder modules. The backbone processes text-aligned semantic representations, creating a ``semantic bottleneck'' where fine-grained acoustic details are compressed or lost. While effective for instruction following, these models cannot natively generate audio and often rely on separate modules like vocoders with fixed speaker embeddings, limiting their utility as end-to-end audio foundation models.

\textbf{Semantic-Only Models.} Approaches like TWIST \cite{twist}, SpiritLM \cite{spiritlm}, VoxtLM \cite{voxtlm}, SUTLM \cite{sutlm}, and SIMS \cite{maimon2025scaling} operate on discrete speech tokens but restrict themselves to semantic tokens (e.g., HuBERT units). VoxtLM and SUTLM combine text BPEs with HuBERT tokens and support ASR, TTS, and continuation via control tokens, but still lack acoustic detail. While SpiritLM introduces token interleaving, it focuses on semantic content, discarding the acoustic details required for acoustic understanding and high-fidelity audio generation.

\textbf{Native Audio Models.} The closest precursors to our work are native models that model discrete acoustic tokens directly. AudioLM \cite{audiolm} pioneered the modeling of semantic and acoustic tokens but relied on a hierarchical cascaded architecture, generating semantic tokens first, followed by acoustic tokens in separate steps. Discrete audio models (such as VALLE \cite{valle}, CosyVoice \cite{du2024cosyvoice}, or Orpheus \cite{orpheus}) have shown success in TTS. Moshi \cite{moshi} introduced a full-duplex model for real-time dialogue, while Llama-Mimi \cite{llamamimi} demonstrated that interleaving semantic and acoustic tokens within a single Llama-3 decoder achieves the best acoustic consistency. However, these works focus on specific speech tasks without systematic investigation of training recipes or scaling behavior---gaps we address.


\subsection{Scaling Laws for Foundation Models}
For LLM pre-training, \citet{kaplan2020} first established power-law relationships suggesting model size should scale faster than data ($N \propto C^{0.73}$). The Chinchilla study \cite{chinchilla} revised this, showing that for compute-optimal text LLMs, model size and training tokens should scale equally ($N^*, D^* \propto C^{0.5}$).

Recent works have attempted to extend scaling laws to the audio domain, yet remain constrained by their focus on \textit{semantic-only} models. Both \citet{cuervo-marxer-2024-scaling} and \citet{maimon2025scaling} limit their analysis to semantic tokens (HuBERT units), effectively discarding the acoustic details required for general speech/audio modeling. While \citet{maimon2025scaling} improves upon the textless approach of \citet{cuervo-marxer-2024-scaling} by demonstrating that interleaving text accelerates learning, their methodology relies on a sparse number of models per compute budget, preventing reliable IsoFLOP curve fitting ($N^*$ vs $D^*$). Furthermore, neither work investigates the scaling behavior of cross-modal capabilities (e.g., ASR and TTS skills), leaving a gap in understanding how audio skills emerge with scale.

\section{Experimental Setup}
\label{sec:experimental_setup}

\begin{figure*}[t]
    \centering
    \includegraphics[width=0.98\textwidth]{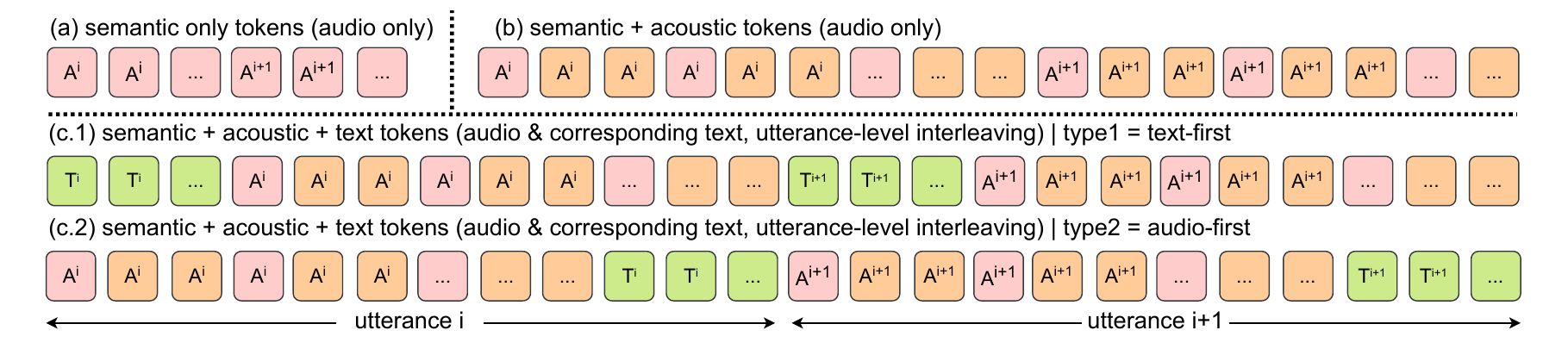}
    \caption{Three token types examined in \S~\ref{sec:token_composition}: (a) Semantic-only, (b) Semantic+Acoustic, and (c) Utterance-level interleaved Semantic+Acoustic+Text, where (c.1) shows text-first format, and (c.2) shows audio-first format. Superscript $i$ denotes utterance index, where utterance($i$) and utterance($i$+1) are adjacent audio segments from the same document. See Appendix~\ref{sec:appendix_formatting} for detailed data formatting.}
    \label{fig:token_types}
\end{figure*}

\textbf{Model Architecture}: We train decoder-only Transformers using the Qwen3 architecture \cite{qwen3}, which adds QK-Norm over Llama for improved training stability, with random initialization (cold-start).\footnote{Section~\ref{sec:warm_cold_comparison} compares cold-start vs. warm-start initialization. Training hyperparameters (optimizer, learning schedule, etc.) and hardware details are in Appendix~\ref{sec:appendix_hyperparameters}.} We discretize audio using Mimi~\cite{moshi} for its high reconstruction quality and semantic-acoustic separation, where the first codebook captures \textit{semantic} content and remaining codebooks capture \textit{acoustic} details. Mimi is a neural codec at 12.5 Hz, and we use first 8 RVQ codebooks (100 tokens/sec). Audio tokens are flattened and interleaved with text at the utterance level (shown in Figure~\ref{fig:token_types}(c)), with both audio-first and text-first variants for each instance. Unlike task-specific models (e.g., Orpheus for TTS \cite{orpheus}, Llama-Mimi for audio continuation \cite{llamamimi}), this interleaving enables the model to be \textit{general-purpose}, learning four capabilities: (1) audio continuation, (2) text continuation, (3) audio$\rightarrow$text, and (4) text$\rightarrow$audio.

\textbf{Training Data}: For speech data, we select the largest publicly available corpora with utterance-level transcriptions: (1) \textit{Yodas}~\cite{li2023yodas} (500K+ hours across 100+ languages; we use $\sim$165K hours of English), (2) \textit{Emilia}~\cite{he2024emilia} (101K hours of diverse, spontaneous speech; we use $\sim$140K hours of English), and (3) \textit{MLS}~\cite{mls} (45K hours of audiobook speech). For text-only data, we use \textit{Nemotron-CC}~\cite{su-etal-2025-nemotron}, a large-scale web corpus widely used in LLM pre-training. Section~\ref{sec:design_choice} ablates these choices; full data statistics are in Appendix~\ref{sec:appendix_data_statistics}.

\textbf{Evaluation}: We evaluate across four categories: (1) \textit{Speech semantic knowledge} (sBLIMP, sWUGGY), (2) \textit{Speech acoustic knowledge} (Salmon), (3) \textit{Text knowledge} (tBLIMP, tWUGGY, HellaSwag), and (4) \textit{Cross-modal skills} (LibriSpeech for ASR, seed-tts-eval for TTS). Full details on evaluation are provided in Appendix~\ref{sec:appendix_evaluation}.

\textbf{SODA-Preliminary}: As an initial exploration, we first validate whether training a vanilla transformer on our utterance-level interleaved discrete audio tokens yields meaningful capabilities. We train SODA-prelim, a 600M parameter model on 500B multilingual tokens (8 languages from Yodas), using multilingual data to maximize available training data from Yodas before processing other large corpora. The results confirm that joint semantic-acoustic-text modeling produces functional cross-modal skills (ASR, TTS) and strong acoustic understanding, but reveals limitations in semantic understanding and text knowledge (see Appendix~\ref{sec:appendix_prelim} for full details). Since standard benchmarks focus on English skills, subsequent experiments (\S\ref{sec:design_choice}) adopt English-only data to isolate multilingual effects.
\section{What Data and Token Types Do We Use?}
\label{sec:design_choice}

Before conducting scaling analysis, we perform empirical investigations to answer: \textit{What speech corpora work best? How much text-only data should we include? What token composition (semantic, acoustic, text) is optimal?} These studies ensure our scaling laws reflect a well-optimized setting rather than a sub-optimal baseline. Each subsection uses a different experimental setup tailored to the question (see Appendix~\ref{sec:appendix_design_choice_setup} for rationale). To isolate data source effects, all experiments use English-only data.

\subsection{What Speech Data Works the Best?}
\label{sec:speech_data_selection}

\textbf{Question}: Which speech data should be used for training? Building on SODA-prelim (\S\ref{sec:experimental_setup}), we compare available corpora to identify a better training recipe. 

\textbf{Setup}: Comparing data sources via full training runs is too expensive. Recent work showed that \textit{annealing experiments}, where different data mixtures are evaluated during the learning rate decay phase of pre-training, can reliably predict the relative quality of data sources at a fraction of the cost~\cite{blakeney2024spark}. We adopt this approach: we branch from the stable phase of SODA-prelim to compare speech datasets during annealing. We evaluate the three speech corpora introduced in \S\ref{sec:experimental_setup}: Yodas, Emilia, and MLS

\textbf{Findings}: Full results are in Table~\ref{tab:speech_data_selection} (Appendix~\ref{sec:appendix_speech_data}); we summarize key findings here. Across semantic and acoustic tasks, all three corpora yield similar results. However, MLS shows notably poor cross-modal performance despite being a curated audiobook corpus: ASR-WER degrades to 92.6\% and TTS-WER to 35.7\%. We attribute this to: (i) uncased, unpunctuated transcripts creating a mismatch with standard text, and (ii) fixed-length 10--20 second chunks lacking length diversity. Between Emilia and Yodas, they show complementary strengths: Emilia achieves the best TTS results, while Yodas provides better text knowledge. Separately, we find that small-scale models trained from scratch (150M, 10B tokens), in line with \citet{datadecide}, confirm this result. Thus, we select \textbf{Yodas + Emilia}.

\subsection{How Much Text-Only Data Should We Include?}
\label{sec:text_data_mixture_ratio}

\textbf{Question}: Given poor semantic understanding and general knowledge observed in SODA-prelim (see Appendix~\ref{sec:appendix_prelim}), can we boost these capabilities by letting the model learn from high-quality text data in addition to noisy speech transcripts during pre-training? If so, what is the optimal ratio?

\textbf{Setup}: Since small-scale runs (150M, 10B tokens) yielded similar findings to expensive annealing in \S\ref{sec:speech_data_selection}, we adopt this setup for the text ratio sweep. We vary the ratio from 0\% to 50\%, where X\% text means sampling X\% tokens from Nemotron and (100$-$X)\% from Yodas.

\begin{figure}[h!] 
    \centering
    \includegraphics[width=0.975\columnwidth]{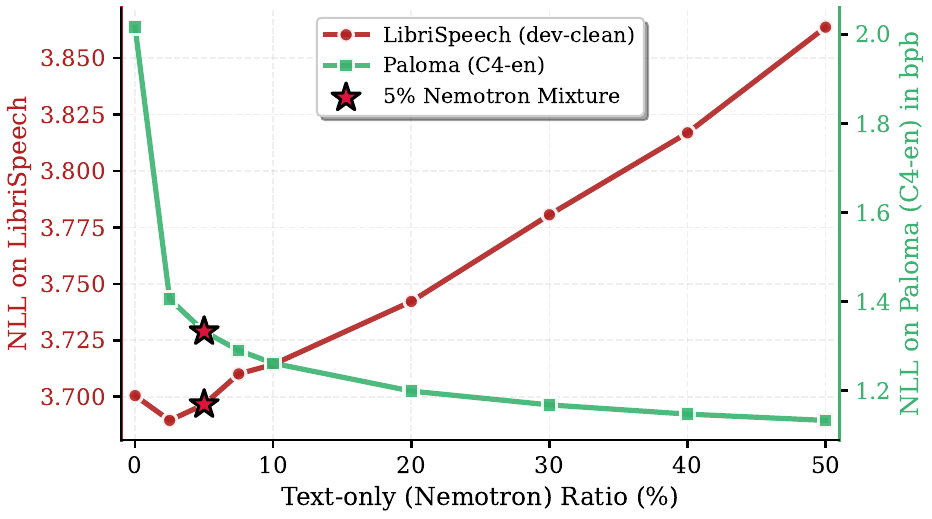}
    \caption{Impact of adding Nemotron on NLL for audio and text validation data. Full results on other metrics are shown in Figure~\ref{fig:nemotron_sweep_appendix}.}
    \label{fig:nemotron_sweep}
\end{figure}

\begin{figure*}[t]
    \centering
    \includegraphics[width=0.99\textwidth]{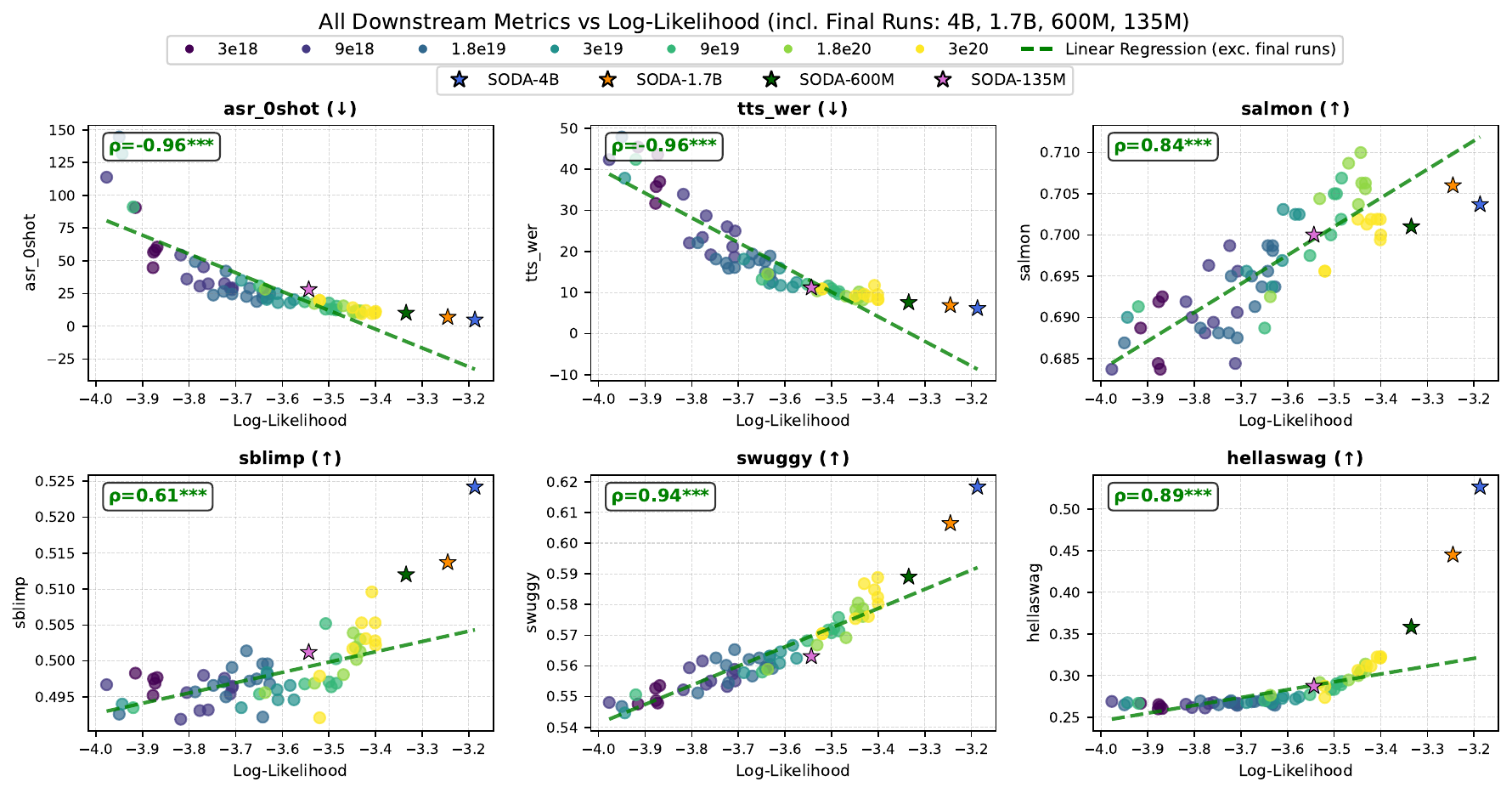}
    \caption{Validation NLL (audio+text) versus downstream task performance. Circular points: 64 IsoFLOP models (\S\ref{sec:scaling}); star-shaped points: final SODA runs (\S\ref{sec:largescale}). Regression lines are fitted on 64 IsoFLOP models only. Full results with other metrics are shown in Appendix~\ref{sec:appendix_nll}.}
    \label{fig:nll_correlation}
\end{figure*}

\textbf{Findings}: Figure~\ref{fig:nemotron_sweep} reveals a trade-off between audio and text performance (we validate NLL as a reliable downstream predictor in \S\ref{sec:validation_loss_metric}). For NLL$_{\text{text}}$, adding any text yields a large improvement (drastic drop in NLL$_{\text{text}}$ from 0\% to 2.5\%), with continued improvement as text ratio increases. NLL$_{\text{audio}}$ does not degrade up to 5\% text (matching the 0\% baseline), but degrades beyond this point. We select \textbf{5\%} as it maximizes text knowledge gains \textit{with little to no degradation} in audio performance---practitioners prioritizing text/reasoning capabilities may increase this ratio at the cost of audio skills. Full results on downstream metrics are in Figure~\ref{fig:nemotron_sweep_appendix} (Appendix~\ref{sec:appendix_nemotron}). We fix the pre-training mixture to \textbf{5\% Text (Nemotron) + 95\% Speech (Yodas/Emilia)} for all subsequent experiments.

\subsection{What Is a Good Token Composition?}
\label{sec:token_composition}

\textbf{Question}: What is the impact of interleaved audio and text tokens, as well as the combination of semantic and acoustic audio tokens, on audio benchmarks?

\textbf{Setup}: We ablate three token types (illustrated in Figure~\ref{fig:token_types}) using a fixed budget of $3 \times 10^{20}$ FLOPs (1.7B model, 30B data) and Yodas as speech data (without Nemotron). The three token configurations are: (1) \textit{Semantic-only}, using only the first Mimi codebook, (2) \textit{Semantic+Acoustic}, using up to 8 Mimi codebooks (audio only), and (3) \textit{Semantic+Acoustic+Text}, interleave text (transcript) with audio tokens at the utterance level.

\begin{table}[h]
    \centering
    \caption{Token ablation results. \texttt{S}=Semantic, \texttt{A}=Acoustic, \texttt{T}=Text. $\times$ indicates the model lacks the capability to perform the task.}
    \label{tab:token_composition}
    \footnotesize
    \setlength{\tabcolsep}{0.85pt}
    \begin{tabular}{@{}lccccccc@{}}
    \toprule
    & \multicolumn{2}{c}{\textbf{Semantic}} & \textbf{Acou.} & \multicolumn{2}{c}{\textbf{Text}} & \multicolumn{2}{c}{\textbf{Cross-Modal}} \\
    \cmidrule(lr){2-3} \cmidrule(lr){4-4} \cmidrule(lr){5-6} \cmidrule(lr){7-8}
    & sBLI$\uparrow$ & sWUG$\uparrow$ & Salm$\uparrow$ & tBLI$\uparrow$ & tWUG$\uparrow$ & ASR$_{\scriptsize\text{WER}}$$\downarrow$ & TTS$_{\scriptsize\text{WER}}$$\downarrow$ \\
    \midrule
    \texttt{S}     & 58.6 & 72.1 & 67.3 & $\times$ & $\times$ & $\times$ & $\times$ \\
    \texttt{S+A}   & 50.9 & 59.0 & 70.1 & $\times$ & $\times$ & $\times$ & $\times$ \\
    \texttt{S+A+T} & 50.4 & 58.1 & 70.4 & 67.8 & 71.6 & 18.3 & 27.1 \\
    \bottomrule
    \end{tabular}
\end{table}

\textbf{Findings}: Table~\ref{tab:token_composition} reveals a trade-off: adding acoustic tokens improves acoustic modeling (Salmon: 67.3\% $\rightarrow$ 70.1\%) but reduces semantic understanding (sBLIMP: 58.6\% $\rightarrow$ 50.9\%). Interleaving text tokens has minimal further impact on these metrics while unlocking cross-modal (ASR/TTS) and text capabilities unavailable in audio-only models like our \texttt{S+A} model variant or Llama-Mimi.

While semantic-only models excel at semantic understanding, they lack the acoustic detail needed for high-fidelity understanding and generation. Since our goal is a \textit{general-purpose} backbone, we adopt \texttt{S+A+T}, accepting the semantic trade-off for broader capabilities in a unified backbone.

\section{How Should We Allocate Compute?}
\label{sec:scaling}

This section addresses: \textit{Is validation loss a reliable metric for discrete audio models? How should we allocate compute between model size and training data?} Scaling laws for text LLMs are well-studied~\cite{kaplan2020,chinchilla,llama3,deepseek}, yet no such analysis exists for discrete audio models. We conduct the first scaling law study for discrete audio models, investigating the extent to which the lower information density of audio tokens (100 tokens/sec vs.\ $\sim$4 tokens/sec for text) alters the compute-optimal allocation between model size $N$ and training data $D$. To determine the compute-optimal allocation, we conduct an IsoFLOP sweep training 64 models across seven compute budgets ranging from $3 \times 10^{18}$ to $3 \times 10^{20}$ FLOPs. For each budget $C$, we train models of varying sizes (from 77M to 4.2B parameters), adjusting the dataset size $D$ to satisfy $C \approx 6ND$, with hyperparameters following existing scaling law work by~\citet{held2025relative}.

\subsection{Is Validation Loss a Reliable Metric?}
\label{sec:validation_loss_metric}

\textbf{Question}: Before conducting IsoFLOP analysis, we assess whether validation loss (NLL) on held-out audio data is a reliable metric for evaluating discrete audio models. If this is the case, next experiments can focus on minimizing NLL.

\textbf{Setup}: We compute NLL on speech utterances from LibriSpeech dev-clean and analyze the correlation between NLL and downstream task performance across 64 models of varying sizes and training configurations. Results in Figure~\ref{fig:nll_correlation}, with full results in Figures~\ref{fig:nll_correlation_full1} and~\ref{fig:nll_correlation_full2} (Appendix~\ref{sec:appendix_nll}).

\textbf{Findings}: Across all compute budgets and model sizes, validation NLL shows \textbf{strong rank correlation (Spearman $\rho$) with downstream performance}. 

For \textit{cross-modal skills}, NLL is highly predictive of both ASR ($\rho \approx 0.95$) and TTS quality (TTS-WER: $\rho \approx 0.96$, TTS-SIM: $\rho = 0.99$ with near-linear improvement). However, at lower loss values, improvement slows as most results from the highest compute budget ($3 \times 10^{20}$ FLOPs) stay worse than the regression line.

For \textit{semantic and acoustic understanding} (Salmon, sBLIMP, sWUGGY), the rate of improvement is slow: as NLL decreases from 4.0 to 3.4, Salmon improves from 68.5\% to 70.5\%, and sWUGGY from 54\% to 58\%. For Salmon, we observe early signs of saturation: models at the highest compute budget ($3 \times 10^{20}$ FLOPs) mostly fall below the regression line, suggesting that acoustic understanding may be reaching a diminishing return point. For sBLIMP, performance remains largely pre-emergence (49.3\% to 50.0\%, near the 50\% random baseline), suggesting this metric is not yet informative at our current scale.

For \textit{text knowledge} tasks (tBLIMP, tWUGGY, HellaSwag; see Figure~\ref{fig:nll_correlation_full2}), NLL correlates strongly ($\rho > 0.8$) and improvement is more pronounced than speech understanding: as NLL decreases from 4.0 to 3.4, tWUGGY improves from 62\% to 69\% and tBLIMP from 64\% to nearly 70\%, with similarly no sign of saturation yet. HellaSwag ($\rho = 0.89$) shows an emergence pattern: no improvement from NLL 4.0 to 3.6, but rapid increase from 25\% to 32\% as NLL drops from 3.6 to 3.4.

\textbf{Choice of NLL Metric}: Our interleaved format admits multiple ways to compute validation NLL (e.g., all tokens from audio-first data, audio-only tokens, text-only tokens). While different variants yield similar correlations, we select \textbf{NLL on all tokens from audio+text data} as our primary metric because it is the simplest (standard NLL over all tokens) and achieves the best balance between speech and text task correlations (see Table~\ref{tab:nll_spearman} in Appendix~\ref{sec:appendix_nll}).

\textbf{Extrapolation to Larger Scale}: We validate whether these NLL--performance trends hold beyond the IsoFLOP regime by including the final SODA runs (colored points in Figure~\ref{fig:nll_correlation}). These models, trained at higher compute budgets (\S\ref{sec:largescale}), largely follow the extrapolated regression lines---confirming NLL as a reliable proxy even at scale. We discuss task-specific extrapolation patterns in detail in \S\ref{sec:largescale}.

\subsection{What Is Compute-Optimal for Discrete Audio?}
\label{sec:isoflop_analysis}
\begin{figure*}[t]
    \centering
    \begin{subfigure}[t]{0.305\textwidth}
        \centering
        \includegraphics[width=\textwidth]{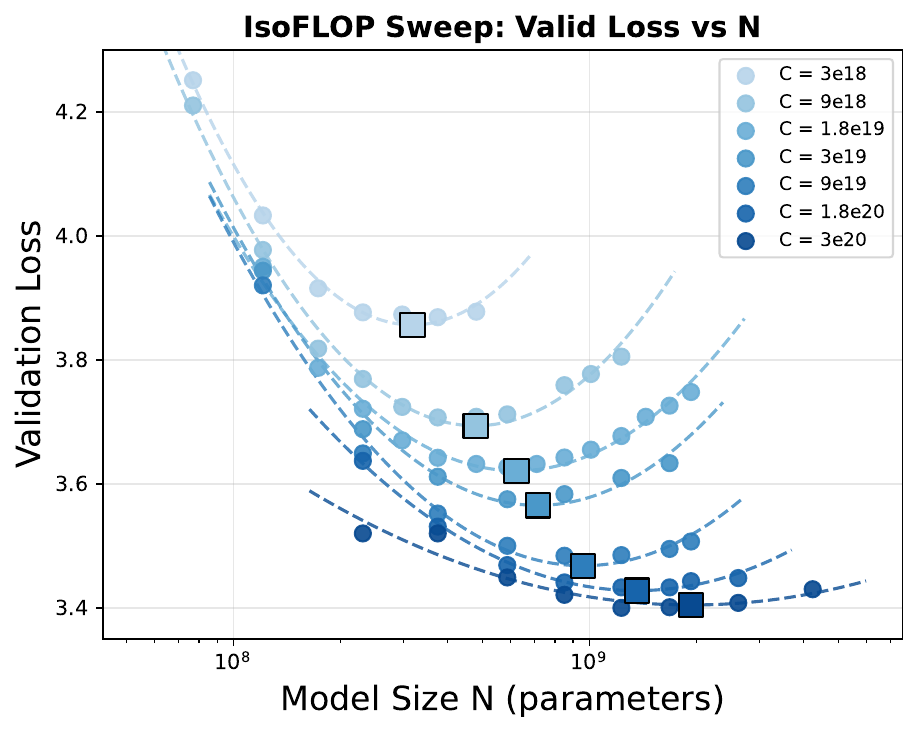}
        \caption{Validation NLL vs model size $N$}
        \label{fig:isoflop_N}
    \end{subfigure}
    \hfill
    \begin{subfigure}[t]{0.305\textwidth}
        \centering
        \includegraphics[width=\textwidth]{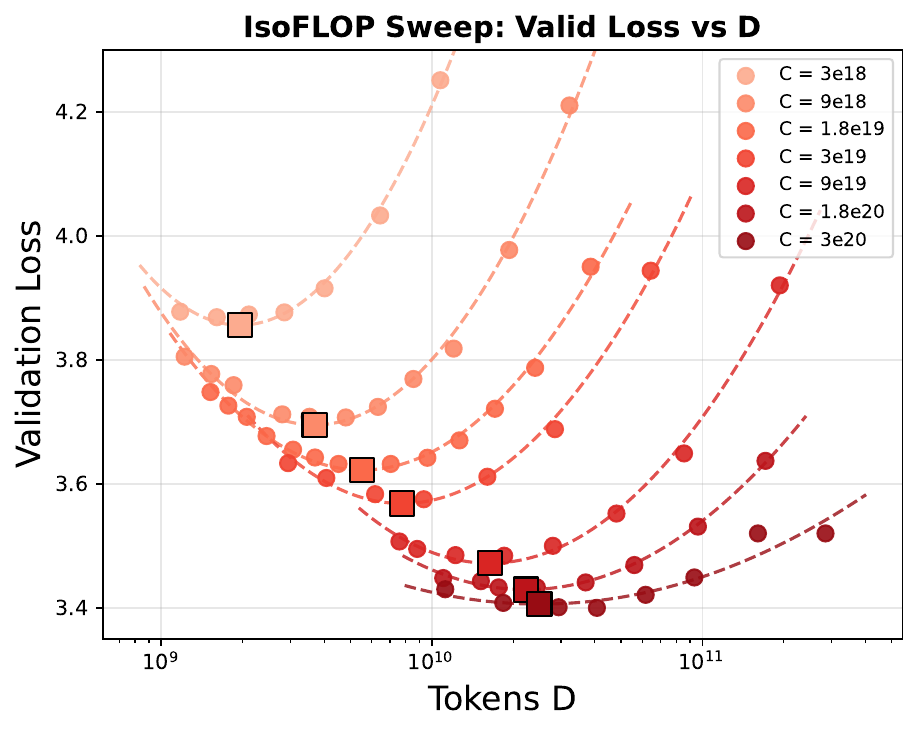}
        \caption{Validation NLL vs training tokens $D$}
        \label{fig:isoflop_D}
    \end{subfigure}
    \hfill
    \begin{subfigure}[t]{0.372\textwidth}
        \centering
        \includegraphics[width=\textwidth]{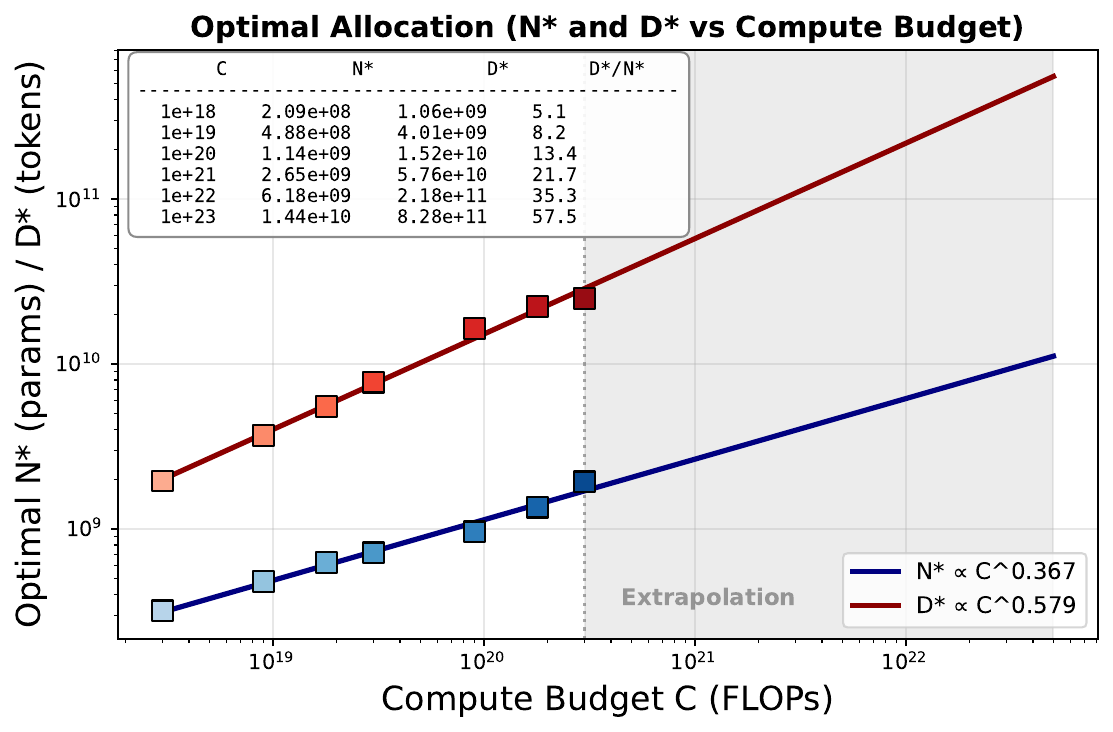}
        \caption{Optimal $N^*$ and $D^*$ vs compute budget $C$}
        \label{fig:isoflop_extrapolation}
    \end{subfigure}
    \caption{IsoFLOP analysis for discrete audio modeling. (a) and (b) show the loss landscape across model sizes and token counts for each compute budget. (c) shows the fitted scaling laws with extrapolation, revealing that optimal data $D^*$ scales faster than optimal model size.}
    \label{fig:isoflop}
\end{figure*}

\textbf{Question}: Having established that validation loss is a reliable metric, how should we allocate compute between model size ($N$) and training data ($D$) to minimize the loss?

\textbf{Setup}: Figure~\ref{fig:isoflop_N} shows validation loss versus model size $N$ for each compute budget. Following Chinchilla's methodology~\cite{chinchilla}, we fit a quadratic function $L = a(\log N)^2 + b(\log N) + c$ to the observed points and identify the compute-optimal model size $N^*$ at the minimum of this curve. Similarly, Figure~\ref{fig:isoflop_D} shows validation loss versus training tokens $D$, with the compute-optimal token count $D^*$ identified analogously. 

\textbf{Findings}.\footnote{Another approach to derive scaling law exponents is to fit a parametric equation $L = E + A/N^\alpha + B/D^\beta$ on all data points and derive exponents as $N^* \propto C^{\beta/(\alpha+\beta)}$ and $D^* \propto C^{\alpha/(\alpha+\beta)}$. This yields similar findings; see Appendix~\ref{sec:appendix_parametric} for details.} Given the compute-optimal $(N^*, D^*)$ identified at each of the seven compute budgets, we fit power-laws $N^* = a_N C^{b_N}$ and $D^* = a_D C^{b_D}$ using log-linear regression to derive scaling laws for discrete audio (illustrated in Figure~\ref{fig:isoflop_extrapolation}). This yields:
\begin{align}
    N^* &\propto C^{0.367} \label{eq:scaling_law_N} \\
    D^* &\propto C^{0.579} \label{eq:scaling_law_D} 
\end{align}
These exponents differ from Chinchilla's scaling laws ($N^*, D^* \propto C^{0.5}$), with data scaling faster than model size. However, recent text LLM studies have found similar trends: Llama3~\cite{llama3} reports an exponent of data scaling of 0.53, and DeepSeek~\cite{deepseek} finds exponents ranging from 0.42 to 0.55 depending on data quality. This DeepSeek work interprets higher data exponents as indicative of lower information density in training data. Our exponent ($D^* \propto C^{0.579}$) is in line with this interpretation, suggesting that discrete audio tokens at 100 tokens/sec carry less information per token than text, requiring more data to achieve the same effective learning. We also note that IsoFLOP curves tend to flatten at larger scales~\cite{llama3}, so our smaller-scale regime ($\leq 3 \times 10^{20}$ FLOPs) may contribute to the higher exponent. To our knowledge, this represents the first scaling analysis for discrete audio models, establishing a foundation for future investigations into how exponents vary with tokenizer design and token rate.\footnote{Studying how scaling exponents change with token rate (reflecting information density) and audio data quality is an interesting direction we leave to future work.}

\textbf{Optimal Token-to-Parameter Ratio}. This asymmetry implies that the optimal token-to-parameter ratio ($D^*/N^*$) is not constant but \textit{increases} with scale. Our projections suggest an optimal ratio of around $13$ tokens per parameter at $10^{20}$ FLOPs, rising to around $58$ tokens per parameter at $10^{23}$ FLOPs. While the Chinchilla study suggested a constant ratio of $20$ tokens per parameter, a replication attempt~\cite{besiroglu2024chinchilla} found this estimate was poorly fit and that the ratio actually \textit{decreases} with compute for text models. Our finding of an \textit{increasing} ratio for discrete audio represents a qualitatively different scaling behavior.


\section{Does Scaling Up Work?}
\label{sec:largescale}

This section addresses: \textit{Does scaling up lead to competitive performance? Should we train from scratch or from text LLMs?} Guided by recipes from \S\ref{sec:design_choice} and scaling analysis from \S\ref{sec:scaling}, we train \textbf{SODA} ({S}caling {O}pen {D}iscrete {A}udio), ranging from 135M to 4B parameters, and evaluate against our scaling predictions and existing spoken language models.

\subsection{Setup and Over-Training}

\textbf{Training data}: Following the recipe from \S\ref{sec:design_choice}, we train on 95\% speech corpora (Yodas + Emilia) and 5\% Nemotron text corpus (\S\ref{sec:experimental_setup}). This yields around 125B + 125B = 250B tokens of interleaved speech data (audio-first and text-first formats), for a total of \textbf{500B tokens} ($\sim$4 epochs). Prior work shows repeated data remains effective up to 4 epochs for LLM pre-training~\cite{muennighoff2023scaling}. 

\textbf{Model sizes and over-training}: We train models at 135M, 600M, 1.7B, and 4B parameters, all on 500B tokens. While our scaling laws define compute-optimal token counts, inference usage favors models trained beyond $D^*$, especially given that our 100 tokens/sec rate can make larger models slow. This results in varying over-training factors: 135M at $\sim$940$\times$ $D^*$, 600M at $\sim$90$\times$ (similar to Llama3), 1.7B at $\sim$18$\times$ (similar to Llama2), and 4B at $\sim$4.5$\times$ (approaching compute-optimal). Training the 4B model on 500B tokens reaches $1.3 \times 10^{22}$ FLOPs ($\sim$1 week on v5p-256 TPU). See Appendix~\ref{sec:appendix_parametric} for detailed over-training analysis.

\subsection{How Well Does SODA Perform?}

\begin{table*}[t]
    \caption{Results comparing SODA against existing Speech Language Models. SODA is the only model with capabilities across all skills, serving as a unified backbone. $\times$ indicates lacks the capability; -- indicates results not reported due to HellaSwag contamination known in Llama2 pre-training which SpiritLM is based on \cite{touvron2023llama}. $^\dagger$SpiritLM uses 10-shot evaluation on LibriSpeech-clean for ASR/TTS whereas ours is 0-shot with TTS on seed-tts-eval; $^\ddagger$TTS-SIM reproduced---semantic-only tokens cannot preserve voice.}
    \label{tab:main_table}
    \centering
    \small
    \setlength{\tabcolsep}{3pt}
    \renewcommand{\arraystretch}{0.975}
    \begin{tabular}{@{}lcccccccccc@{}}
    \toprule
    & \multicolumn{2}{c}{\textbf{Speech (Semantic)}} & \textbf{Speech (Acoustic)} & \multicolumn{3}{c}{\textbf{Text (Knowledge)}} & \multicolumn{3}{c}{\textbf{Cross-Modal}} \\
    \cmidrule(lr){2-3} \cmidrule(lr){4-4} \cmidrule(lr){5-7} \cmidrule(lr){8-10}
    \textbf{Model} & \textbf{sBLIMP}$\uparrow$ & \textbf{sWUGGY}$\uparrow$ & \textbf{Salmon}$\uparrow$ & \textbf{tBLIMP}$\uparrow$ & \textbf{tWUGGY}$\uparrow$ & \textbf{HellaS}$\uparrow$ & \textbf{ASR}$\downarrow$ & \textbf{TTS$_{\text{WER}}$}$\downarrow$ & \textbf{TTS$_{\text{SIM}}$}$\uparrow$ \\
    \midrule
    TWIST-7B & 59.0 & 73.9 & 61.6 & $\times$ & $\times$ & $\times$ & $\times$ & $\times$ & $\times$ \\ 
    SpiritLM-base-7B & 58.3 & 69.0 & 57.2 & 73.3 & 80.3 & -- & $\sim$22$^\dagger$ & $\sim$40$^\dagger$ & $\sim$0.05$^\ddagger$ \\
    SpiritLM-Expr-7B & 54.2 & 65.0 & 67.1 & 73.6 & 75.8 & -- & $\sim$38$^\dagger$ & $\sim$50$^\dagger$ & $\sim$0.10$^\ddagger$ \\
    Llama-Mimi-1.3B & 54.3 & 68.7 & 73.6 & $\times$ & $\times$ & $\times$ & $\times$ & $\times$ & $\times$ \\
    Llama-Mimi-8B & 55.1 & 68.8 & 73.2 & $\times$ & $\times$ & $\times$ & $\times$ & $\times$ & $\times$ \\
    \midrule
    SODA-prelim-600M & 50.9 & 57.8 & 69.4 & 69.0 & 71.3 & 26.2 & 22.0 & 9.2 & 0.516 \\
    \midrule
    SODA-base-135M & 50.1 & 56.3 & 70.0 & 67.4 & 70.7 & 28.7 & 28.1 & 11.2 & 0.500 \\
    SODA-base-600M & 51.2 & 58.9 & 70.1 & 70.7 & 73.1 & 35.8 & 10.2 & 7.6 & 0.555 \\
    SODA-base-1.7B & 51.4 & 60.6 & 70.6 & 70.3 & 74.7 & 44.5 & 7.0 & 6.9 & 0.560 \\
    SODA-base-4B & 52.4 & 61.8 & 70.4 & 71.3 & 74.8 & 52.6 & 5.0 & 6.1 & 0.560 \\
    \bottomrule
    \end{tabular}
\end{table*}

\subsubsection{Do Our Design Choices Help?}
Comparing SODA-600M-base to SODA-600M-prelim (both 600M parameters trained on 500B tokens) validates that our design choices in \S\ref{sec:design_choice}, including the shift to English-only and data mixture, yield measurable improvements. Table~\ref{tab:main_table} shows that SODA-600M-base improves across all metrics: notably, ASR-WER drops from 22.0\% to 10.2\%, TTS-WER from 9.2\% to 7.6\%, and TTS-SIM increases from 0.516 to 0.555. Semantic understanding (sWUGGY: 57.8\% $\rightarrow$ 58.9\%) and text knowledge (tWUGGY: 71.3\% $\rightarrow$ 73.1\%) also improve. These gains across all skills reflect the refined recipe established in \S\ref{sec:design_choice}.

\subsubsection{Does Performance Improve with Scale?}

We examine how downstream task performance improves with scale, connecting the NLL correlation analysis from \S\ref{sec:validation_loss_metric} to our final SODA runs (Table~\ref{tab:main_table} and Figure~\ref{fig:nll_correlation}). For validation loss predictions under over-training, see Appendix~\ref{sec:appendix_parametric}.

\textbf{Cross-modal tasks} (ASR, TTS) showed strong NLL correlation ($\rho > 0.95$) with early saturation signs in Figure~\ref{fig:nll_correlation}. The final runs confirm this: ASR-WER drops dramatically from 28.1\% (135M) to 5.0\% (4B), but all final-run points fall above the regression line, indicating diminishing returns at scale. TTS evaluation follows a similar pattern.

\textbf{Acoustic understanding} (Salmon), which showed saturation at the highest compute budgets, plateaus around 70\% across all model sizes (70.0\% at 135M to 70.4\% at 4B). Final runs fall below the regression line in Figure~\ref{fig:nll_correlation}, confirming saturation---acoustic ability appears bounded by tokenization and/or data quality rather than model capacity.

\textbf{Semantic understanding} shows emergence. While sWUGGY and sBLIMP showed slow improvement in IsoFLOP models in Figure~\ref{fig:nll_correlation}, final SODA runs fall \textit{above} the regression line: sWUGGY improves from 56.3\% (135M) to 61.8\% (4B), and sBLIMP from 50.1\% to 52.4\%, suggesting emergence and accelerating gains at larger scales.

\textbf{Text knowledge} exhibits the strongest emergence. Our analysis in Fig.~\ref{fig:nll_correlation} hinted at emergence (rapid improvement as NLL $<$ 3.6), and the final runs confirm exponential gains: accuracy increases from 28.7\% (135M) to 52.6\% (4B), with final-run points appearing far above the regression line.


\subsubsection{How Does SODA Compare to Others?}

\textit{SpiritLM} uses semantic-only tokens (HuBERT), achieving strong semantic understanding (sBLIMP: 58.3\%, sWUGGY: 69.0\%) but weaker acoustic modeling (Salmon: 57.2--67.1\%). While SODA's semantic scores are lower with our interleaved setup, we show in Table~\ref{tab:token_composition} (\S\ref{sec:token_composition}) that a semantic-only SODA variant at $3 \times 10^{20}$ FLOPs achieves sBLIMP: 58.6\% and sWUGGY: 72.1\%, outperforming both SpiritLM variants. This confirms that the semantic-acoustic trade-off is a design choice, not a fundamental limitation. However, semantic-only models—whether SpiritLM or our semantic-only variant—are limited in practice: they primarily model \textit{what} is said but not \textit{how} it is said, lacking the acoustic details necessary for general audio capabilities, reflected in extremely low TTS-SIM scores in Table~\ref{tab:main_table}.

\textit{Llama-Mimi} achieves the highest acoustic scores (Salmon: 73.6\%) by modeling audio-only sequences. Although SODA achieves lower acoustic scores, our ablation in Table~\ref{tab:token_composition} shows that incorporating interleaved text tokens does not degrade this capability (Salmon: 70.1\% $\rightarrow$ 70.4\%), suggesting the gap is attributable to training data differences rather than model design. Nevertheless, Llama-Mimi lacks cross-modal capabilities entirely as it cannot perform ASR or TTS. Our utterance-level interleaving makes SODA a more general-purpose foundation, capable of both audio understanding and speech$\leftrightarrow$text tasks in a unified model.

\subsection{Should We Train From Scratch or Warm-Start?}
\label{sec:warm_cold_comparison}

\textbf{Question}: Many discrete audio models (e.g., TWIST, CSM) initialize from pre-trained text LLMs. Does this \textit{warm-start} strategy benefit over training from scratch (\textit{cold-start})?

\textbf{Setup}: We compare warm-start (from Qwen3-0.6B/1.7B-base) versus cold-start at 600M and 1.7B scales, training on 500B tokens. We provide final evaluation results as well as throughout training trajectories, comparing warm-start vs. cold-start in Table~\ref{tab:warm_cold_comparison} and Figure~\ref{fig:warm_cold_comparison} in Appendix~\ref{sec:appendix_warmstart}.


\textit{Note on Training Stability}: Warm-Start exhibits \textit{instability}, with unpredictable loss spikes (including a big spike at 135K steps in 600M, ASR degrades from 21\% to 34\%; see Figure~\ref{fig:train_loss_stability}). Cold-Start shows smooth improvement throughout.\looseness=-1

\textbf{Cross-Modal Skills}. Cold-Start outperforms Warm-Start on ASR from the earliest checkpoint and maintains this advantage: at 1.7B, Cold-Start achieves 19.7\% vs 29.2\% WER at 10K steps, widening to 7.0\% vs 17.3\% at completion. This suggests warm-start may interfere with learning audio$\rightarrow$text mappings. For TTS, Cold-Start is initially better but Warm-Start catches up; both achieve comparable quality at completion ($\sim$6.5--7.5\% WER, $\sim$0.56 SIM).

\textbf{Speech Understanding.} Salmon ($\sim$70\%) and sWUGGY ($\sim$60\%) show similar trajectories for both starts, suggesting these audio skills are learned regardless of initialization.

\textbf{Text Knowledge.} Warm-Start begins with a substantial advantage (at 10K steps: tWUGGY 75.4\% vs 69.6\%, HellaSwag 40.5\% vs 29.9\%). Critically, Cold-Start \textit{never} catches up even after 500B tokens (final tWUGGY: 74.7\% vs 79.2\%; HellaSwag: 44.5\% vs 47.1\%), indicating text knowledge from LLM pre-training is not fully recoverable through audio-centric training.

\textbf{Recommendation.} Given the training instability and ASR degradation, we recommend \textbf{Cold-Start} as the default recipe for general audio capabilities. However, for capabilities where complex reasoning or knowledge is required, the text-knowledge advantage of Warm-Start may outweigh this. Future work could also consider hybrid approach, e.g., cold-start pre-training followed by text-enriched fine-tuning.

\subsection{Could a New Audio-to-Audio Task Be Formulated as Next-Token Prediction by Fine-tuning SODA?}
\label{sec:s2st}

\textbf{Question}: Can SODA serve as a flexible backbone for new audio tasks? To test this, we fine-tune for \textit{voice-preserving} speech-to-speech translation (S2ST)---an audio$\rightarrow$audio task where translated speech must retain the source voice.

Prior work uses specialized architectures: Translatotron2 \cite{translatotron2} requires a speech encoder, phoneme decoder, spectrogram synthesizer, and shared speaker embeddings. SODA requires none of these---we simply format S2ST as interleaved next-token prediction (source audio $\rightarrow$ source text $\rightarrow$ target text $\rightarrow$ target audio) using the same decoder-only transformer as all other tasks.

\textbf{Setup}: We fine-tune on CVSS-T~\cite{cvss}, a multilingual voice-preserving S2ST corpus (21 languages $\rightarrow$ English) where target speech is synthesized to match the source speaker's voice. We compare 600M models: (1) \textit{SODA}: initialized from SODA-600M (English-only pre-training); (2) \textit{SODA-P}: initialized from SODA-prelim (multilingual pre-training, see \S\ref{sec:experimental_setup}); (3) \textit{Qwen3}: initialized from Qwen3-0.6B; and (4) \textit{Scratch}: random initialization. We evaluate on 200 held-out examples from each language using ASR-BLEU for translation quality and speaker similarity (SIM) for voice preservation. See Appendix~\ref{sec:appendix_s2st} for details.

\begin{table}[h!]
    \centering
    \caption{Speech-to-speech (X$\rightarrow$En) translation on CVSS-T. SODA-Prelim, despite weaker overall recipe and English performance, was pre-trained on multilingual Yodas (en, es, fr, de, th, ar, hi, zh).}
    \label{tab:s2st}
    \small
    \begin{tabular}{@{}l@{\hspace{10pt}}c@{\hspace{10pt}}cccc@{\hspace{10pt}}c@{}}
    \toprule
    & & \multicolumn{4}{c}{\textbf{X$\rightarrow$En (BLEU$\uparrow$)}} & \\
    \cmidrule(lr){3-6}
    \textbf{Init.} & \textbf{Pre-training} & \textbf{es} & \textbf{fr} & \textbf{de} & \textbf{other} & \textbf{SIM}$\uparrow$ \\
    \midrule
    Scratch & None         & 3.5 & 4.5 & 3.5 & 2.7 & 0.349 \\
    Qwen3   & Text         & 3.0 & 4.5 & 3.0 & 2.1 & 0.375 \\ 
    SODA    & Text + Audio & 13.5 & 14.7 & 9.4  & 4.3 & 0.466 \\
    SODA-P  & Text + Audio & 20.9 & 21.1 & 14.6 & 4.5 & 0.469 \\
    \bottomrule
    \end{tabular}
\end{table}

\textbf{Findings}: Table~\ref{tab:s2st} shows that audio pre-training is crucial. SODA outperforms Scratch by 3--4$\times$ on BLEU and achieves substantially better voice preservation (SIM: 0.466 vs.\ 0.349). Text-only pre-training (Qwen3) provides no gains in BLEU over Scratch, confirming that \textit{audio} pre-training---not just scale---drives the improvement. Notably, SODA-P outperforms SODA despite being a weaker backbone on English benchmarks (Table~\ref{tab:main_table}): its multilingual pre-training provides ASR capabilities in the source languages that SODA lacks. This demonstrates that task-specific skills can be enhanced by including relevant training data. For reference, prior work reports SIM scores of $\sim$0.30--0.41 \cite{hibiki}, and ASR-BLEU of $\sim$30\% on es, fr, de \cite{zheng2025rosettaspeech} on related benchmarks, though direct comparison is difficult due to different evaluation protocols.

\textbf{Takeaway}: The same decoder-only transformer and NTP objective used for ASR/TTS, and audio continuation directly supports voice-preserving S2ST---validating SODA as a flexible backbone. Also, individual skills can be enhanced by adding task-specific data without architectural changes.
\section{Conclusion}
\label{sec:conclusion}

Given reasonable data and compute, this work answers: \textit{how to train discrete audio models with NTP to achieve broad capabilities}, providing six findings: (1) adding text-only data improves text knowledge without degrading audio; (2) adding acoustic tokens enables high-fidelity generation but reduces semantic understanding. Future work 
should investigate balancing these trade-offs; (3) optimal data scales 1.6$\times$ faster than model size, consistent with lower information density as in LLM studies; (4) NLL reliably predicts downstream performance, even at larger scales; (5) cold-start outperforms warm-start for audio tasks; (6) capabilities scale differently---cross-modal and acoustic skills saturate, while semantic and text knowledge show accelerating gains.


\section*{Limitations}

\textit{Limited Emergent Capabilities.} Unlike large language models that exhibit emergent abilities at scale, we do not observe strong emergent skills in speech or audio capabilities from pre-training alone. Our evaluation demonstrates the model's utility through fine-tuning on speech-to-speech translation. Future work could study how or when diverse audio capabilities emerge naturally or through audio few-shot learning.

\textit{Generalist vs. Specialist Trade-offs.} As a unified foundation model, SODA prioritizes flexibility over optimizing for any single skill. While specialized models developed for specific tasks may achieve higher performance on their target benchmarks, SODA's strength lies in its ability to formulate audio tasks as next-token prediction using the same underlying model, which could serve as a potential avenue for native audio pre-training.

\textit{Unexplored Design Choices.} Building discrete audio foundation models involves numerous design decisions. While our study investigates what we believe are the primary factors, there exist other design choices that remain unexplored, for example, alternative tokenizers, different token sampling rates, RVQ token-weighted loss functions, and various architectural modifications. Future work could study these additional design choices to further refine training recipes.

\section*{Impact Statement}
\label{sec:impact}


\textit{Positive Impacts.} By releasing open models, training recipes, and scaling laws, this work democratizes access to audio foundation model research. The unified architecture supporting broad speech/audio capabilities could benefit accessibility and the use of audio interfaces in AI applications.

\textit{Potential Risks.} Audio generation capabilities raise concerns about misuse, including voice cloning for impersonation or misleading content generation. High-fidelity speech synthesis could be exploited for deepfakes or fraud.

\textit{Mitigation Considerations.} We encourage users to implement safeguards such as audio watermarking, voice-clone consent requirements, and synthetic speech detection.

\section*{Author Contributions}
Potsawee Manakul and Will Held conceived and led the project. Potsawee Manakul led execution, implemented data processing and training, conducted experiments, and wrote the manuscript. Will Held established the infrastructure and initial training, provided direction, contributed to technical implementation, and provided feedback. Woody Gan assisted with analysis and fine-tuning explorations. Martijn Bartelds contributed to ideation, analysis, and review. Guangzhi Sun provided early direction, implemented evaluation, conducted sanity checks, and reviewed. Diyi Yang is the primary supervisor, providing resources, guidance, direction, and feedback throughout the project.

\section*{Acknowledgements}
We thank the Marin and OpenAthena team for developing and maintaining the training infrastructure. We are particularly grateful to David Hall for his critical support with infrastructure and insights on LLM pre-training. 
This work is supported by ONR N00014-24-1-2532, Sloan Faculty Fellowship to DY, and SCB 10X. 
The compute resource was supported by the Google TPU Research Cloud (TRC) and a Stanford HAI–GCP Grant as part of the Marin Project. PM is supported by SCB 10X for his joint-position at Stanford. We also thank Marin team, SALT Lab members, and Typhoon members for helpful discussions and feedback.


\bibliography{references}

@article{blakeney2024spark,
  title={Does your data spark joy? Performance gains from domain upsampling at the end of training},
  author={Blakeney, Cody and Gururangan, Suchin and Jain, Shrimai},
  journal={arXiv preprint arXiv:2406.03476},
  year={2024}
}

@article{besiroglu2024chinchilla,
  title={Chinchilla scaling: A replication attempt},
  author={Besiroglu, Tamay and Erdil, Ege and Barnett, Matthew and You, Josh},
  journal={arXiv preprint arXiv:2404.10102},
  year={2024}
}

@article{llama3,
  title={The llama 3 herd of models},
  author={Grattafiori, Aaron and Dubey, Abhimanyu and Jauhri, Abhinav and Pandey, Abhinav and Kadian, Abhishek and Al-Dahle, Ahmad and Letman, Aiesha and Mathur, Akhil and Schelten, Alan and Vaughan, Alex and others},
  journal={arXiv preprint arXiv:2407.21783},
  year={2024}
}

@article{touvron2023llama,
  title={Llama 2: Open foundation and fine-tuned chat models},
  author={Touvron, Hugo and Martin, Louis and Stone, Kevin and Albert, Peter and Almahairi, Amjad and Babaei, Yasmine and Bashlykov, Nikolay and Batra, Soumya and Bhargava, Prajjwal and Bhosale, Shruti and others},
  journal={arXiv preprint arXiv:2307.09288},
  year={2023}
}

@article{chinchilla,
  title={Training Compute-Optimal Large Language Models},
  author={Hoffmann, Jordan and Borgeaud, Sebastian and Mensch, Arthur and Buchatskaya, Elena and Cai, Trevor and Rutherford, Eliza and de Las Casas, Diego and Guy, Lisa A and others},
  journal={arXiv preprint arXiv:2203.15556},
  year={2022}
}

@article{kaplan2020,
  title={Scaling Laws for Neural Language Models},
  author={Kaplan, Jared and McCandlish, Sam and Henighan, Tom and Brown, Tom B and Chess, Benjamin and Child, Rewon and Gray, Scott and Radford, Alec and Wu, Jeffrey and Amodei, Dario},
  journal={arXiv preprint arXiv:2001.08361},
  year={2020}
}

@article{deepseek,
  title={Deepseek {LLM}: Scaling open-source language models with longtermism},
  author={Bi, Xiao and Chen, Deli and Chen, Guanting and Chen, Shanhuang and Dai, Damai and Deng, Chengqi and Ding, Honghui and Dong, Kai and Du, Qiushi and Fu, Zhe and others},
  journal={arXiv preprint arXiv:2401.02954},
  year={2024}
}

@inproceedings{maimon2025scaling,
title={Scaling Analysis of Interleaved Speech-Text Language Models},
author={Gallil Maimon and Michael Hassid and Amit Roth and Yossi Adi},
booktitle={Second Conference on Language Modeling},
year={2025},
url={https://openreview.net/forum?id=IXwgE8hyJs}
}

@inproceedings{cuervo-marxer-2024-scaling,
    title = "Scaling Properties of Speech Language Models",
    author = "Cuervo, Santiago  and
      Marxer, Ricard",
    editor = "Al-Onaizan, Yaser  and
      Bansal, Mohit  and
      Chen, Yun-Nung",
    booktitle = "Proceedings of the 2024 Conference on Empirical Methods in Natural Language Processing",
    month = nov,
    year = "2024",
    address = "Miami, Florida, USA",
    publisher = "Association for Computational Linguistics",
    url = "https://aclanthology.org/2024.emnlp-main.21/",
    doi = "10.18653/v1/2024.emnlp-main.21",
    pages = "351--361",
}

@article{llamaomni,
title={{LL}a{MA}-Omni: Seamless Speech Interaction with Large Language Models},
author={Qingkai Fang and Shoutao Guo and Yan Zhou and Zhengrui Ma and Shaolei Zhang and Yang Feng},
booktitle={The Thirteenth International Conference on Learning Representations},
year={2025},
url={https://openreview.net/forum?id=PYmrUQmMEw}
}

@article{moshi,
  title={Moshi: a speech-text foundation model for real-time dialogue},
  author={D{\'e}fossez, Alexandre and Mazar{\'e}, Laurent and Orsini, Manu and Royer, Am{\'e}lie and P{\'e}rez, Patrick and J{\'e}gou, Herv{\'e} and Grave, Edouard and Zeghidour, Neil},
  journal={arXiv preprint arXiv:2410.00037},
  year={2024}
}

@article{llamamimi,
  title={Llama-Mimi: Speech Language Models with Interleaved Semantic and Acoustic Tokens},
  author={Sugiura, Issa and Kurita, Shuhei and Oda, Yusuke and Higashinaka, Ryuichiro},
  journal={arXiv preprint arXiv:2509.14882},
  year={2025}
}

@article{spiritlm,
    title = "{S}pi{R}it-{LM}: Interleaved Spoken and Written Language Model",
    author = "Nguyen, Tu Anh  and
      Muller, Benjamin  and
      Yu, Bokai  and
      Costa-jussa, Marta R.  and
      Elbayad, Maha  and
      Popuri, Sravya  and
      Ropers, Christophe  and
      Duquenne, Paul-Ambroise  and
      Algayres, Robin  and
      Mavlyutov, Ruslan  and
      Gat, Itai  and
      Williamson, Mary  and
      Synnaeve, Gabriel  and
      Pino, Juan  and
      Sagot, Beno{\^i}t  and
      Dupoux, Emmanuel",
    journal = "Transactions of the Association for Computational Linguistics",
    volume = "13",
    year = "2025",
    address = "Cambridge, MA",
    publisher = "MIT Press",
    url = "https://aclanthology.org/2025.tacl-1.2/",
    doi = "10.1162/tacl_a_00728",
    pages = "30--52",
}

@inproceedings{voxtlm,
  title={Voxtlm: Unified decoder-only models for consolidating speech recognition, synthesis and speech, text continuation tasks},
  author={Maiti, Soumi and Peng, Yifan and Choi, Shukjae and Jung, Jee-weon and Chang, Xuankai and Watanabe, Shinji},
  booktitle={ICASSP 2024-2024 IEEE International Conference on Acoustics, Speech and Signal Processing (ICASSP)},
  pages={13326--13330},
  year={2024},
  organization={IEEE}
}

@article{sutlm,
    title = "Toward Joint Language Modeling for Speech Units and Text",
    author = "Chou, Ju-Chieh  and
      Chien, Chung-Ming  and
      Hsu, Wei-Ning  and
      Livescu, Karen  and
      Babu, Arun  and
      Conneau, Alexis  and
      Baevski, Alexei  and
      Auli, Michael",
    editor = "Bouamor, Houda  and
      Pino, Juan  and
      Bali, Kalika",
    booktitle = "Findings of the Association for Computational Linguistics: EMNLP 2023",
    month = dec,
    year = "2023",
    address = "Singapore",
    publisher = "Association for Computational Linguistics",
    url = "https://aclanthology.org/2023.findings-emnlp.438/",
    doi = "10.18653/v1/2023.findings-emnlp.438",
    pages = "6582--6593"
}

@article{qwenomni,
  title={Qwen3-Omni Technical Report},
  author={{Qwen Team}},
  journal={arXiv preprint arXiv:2509.17765},
  year={2025}
}

@inproceedings{salmonn,
title={{SALMONN}: Towards Generic Hearing Abilities for Large Language Models},
author={Changli Tang and Wenyi Yu and Guangzhi Sun and Xianzhao Chen and Tian Tan and Wei Li and Lu Lu and Zejun MA and Chao Zhang},
booktitle={The Twelfth International Conference on Learning Representations},
year={2024},
url={https://openreview.net/forum?id=14rn7HpKVk}
}

@article{twist,
  title={Textually pretrained speech language models},
  author={Hassid, Michael and Remez, Tal and Nguyen, Tu Anh and Gat, Itai and Conneau, Alexis and Kreuk, Felix and Copet, Jade and Defossez, Alexandre and Synnaeve, Gabriel and Dupoux, Emmanuel and others},
  journal={Advances in Neural Information Processing Systems},
  volume={36},
  pages={63483--63501},
  year={2023}
}

@article{valle,
  title={Neural codec language models are zero-shot text to speech synthesizers},
  author={Wang, Chengyi and Chen, Sanyuan and Wu, Yu and Zhang, Ziqiang and Zhou, Long and Liu, Shujie and Chen, Zhuo and Liu, Yanqing and Wang, Huaming and Li, Jinyu and others},
  journal={arXiv preprint arXiv:2301.02111},
  year={2023}
}

@misc{orpheus,
    title={Orpheus-TTS: Human-Like Speech Synthesis},
    author={{Canopy Labs}},
    year={2025},
    howpublished={\url{https://huggingface.co/canopylabs/orpheus-3b-0.1-ft}},
    note={Llama-based TTS model using SNAC tokens}
}

@article{held2025relative,
  title={Relative Scaling Laws for LLMs},
  author={Held, William and Hall, David and Liang, Percy and Yang, Diyi},
  journal={arXiv preprint arXiv:2510.24626},
  year={2025}
}

@article{muennighoff2023scaling,
  title={Scaling data-constrained language models},
  author={Muennighoff, Niklas and Rush, Alexander and Barak, Boaz and Le Scao, Teven and Tazi, Nouamane and Piktus, Aleksandra and Pyysalo, Sampo and Wolf, Thomas and Raffel, Colin A},
  journal={Advances in Neural Information Processing Systems},
  volume={36},
  pages={50358--50376},
  year={2023}
}

@article{chameleon,
  title={Chameleon: Mixed-Modal Early-Fusion Foundation Models},
  author={{Chameleon Team}},
  journal={arXiv preprint arXiv:2405.09818},
  year={2024}
}

@article{qwen3,
  title={Qwen3 technical report},
  author={{Qwen Team}},
  journal={arXiv preprint arXiv:2505.09388},
  year={2025}
}

@inproceedings{li2023yodas,
  title={Yodas: Youtube-Oriented Dataset for Audio and Speech},
  author={Li, Xinjian and Takamichi, Shinnosuke and Saeki, Takaaki and Chen, William and Shiota, Sayaka and Watanabe, Shinji},
  booktitle={2023 IEEE Automatic Speech Recognition and Understanding Workshop (ASRU)},
  pages={1--8},
  year={2023},
  organization={IEEE}
}

@inproceedings{he2024emilia,
  title={Emilia: An extensive, multilingual, and diverse speech dataset for large-scale speech generation},
  author={He, Haorui and Shang, Zengqiang and Wang, Chaoren and Li, Xuyuan and Gu, Yicheng and Hua, Hua and Liu, Liwei and Yang, Chen and Li, Jiaqi and Shi, Peiyang and others},
  booktitle={2024 IEEE Spoken Language Technology Workshop (SLT)},
  pages={885--890},
  year={2024},
  organization={IEEE}
}

@article{mls,
  title={{MLS}: A Large-Scale Multilingual Dataset for Speech Research},
  author={Vineel Pratap and Qiantong Xu and Anuroop Sriram and Gabriel Synnaeve and Ronan Collobert},
  journal={ArXiv},
  year={2020},
  volume={abs/2012.03411}
}

@article{audiolm,
author = {Borsos, Zal\'{a}n and Marinier, Rapha\"{e}l and Vincent, Damien and Kharitonov, Eugene and Pietquin, Olivier and Sharifi, Matt and Roblek, Dominik and Teboul, Olivier and Grangier, David and Tagliasacchi, Marco and Zeghidour, Neil},
title = {AudioLM: A Language Modeling Approach to Audio Generation},
year = {2023},
issue_date = {2023},
publisher = {IEEE Press},
volume = {31},
issn = {2329-9290},
url = {https://doi.org/10.1109/TASLP.2023.3288409},
doi = {10.1109/TASLP.2023.3288409},
journal = {IEEE/ACM Trans. Audio, Speech and Lang. Proc.},
month = jun,
pages = {2523–2533},
numpages = {11}
}

@article{du2024cosyvoice,
  title={Cosyvoice 2: Scalable streaming speech synthesis with large language models},
  author={Du, Zhihao and Wang, Yuxuan and Chen, Qian and Shi, Xian and Lv, Xiang and Zhao, Tianyu and Gao, Zhifu and Yang, Yexin and Gao, Changfeng and Wang, Hui and others},
  journal={arXiv preprint arXiv:2412.10117},
  year={2024}
}

@inproceedings{cvss,
  title={{CVSS} Corpus and Massively Multilingual Speech-to-Speech Translation},
  author={Jia, Ye and Tadmor Ramanovich, Michelle and Wang, Quan and Zen, Heiga},
  booktitle={Proceedings of the Language Resources and Evaluation Conference (LREC)},
  pages={6691--6703},
  year={2022}
}

@inproceedings{translatotron2,
  title={Translatotron 2: High-quality direct speech-to-speech translation with voice preservation},
  author={Jia, Ye and Tadmor Ramanovich, Michelle and Remez, Tal and Pomerantz, Roi},
  booktitle={International Conference on Machine Learning},
  pages={10120--10134},
  year={2022},
  organization={PMLR}
}

@inproceedings{hu2024minicpm,
  title={Mini{CPM}: Unveiling the Potential of Small Language Models with Scalable Training Strategies},
  author={Shengding Hu and Yuge Tu and Xu Han and Ganqu Cui and Chaoqun He and Weilin Zhao and Xiang Long and Zhi Zheng and Yewei Fang and Yuxiang Huang and Xinrong Zhang and Zhen Leng Thai and Chongyi Wang and Yuan Yao and Chenyang Zhao and Jie Zhou and Jie Cai and Zhongwu Zhai and Ning Ding and Chao Jia and Guoyang Zeng and dahai li and Zhiyuan Liu and Maosong Sun},
  booktitle={First Conference on Language Modeling},
  year={2024},
  url={https://openreview.net/forum?id=3X2L2TFr0f}
}

@inproceedings{su-etal-2025-nemotron,
    title = "Nemotron-{CC}: Transforming {C}ommon {C}rawl into a Refined Long-Horizon Pretraining Dataset",
    author = "Su, Dan  and
      Kong, Kezhi  and
      Lin, Ying  and
      Jennings, Joseph  and
      Norick, Brandon  and
      Kliegl, Markus  and
      Patwary, Mostofa  and
      Shoeybi, Mohammad  and
      Catanzaro, Bryan",
    editor = "Che, Wanxiang  and
      Nabende, Joyce  and
      Shutova, Ekaterina  and
      Pilehvar, Mohammad Taher",
    booktitle = "Proceedings of the 63rd Annual Meeting of the Association for Computational Linguistics (Volume 1: Long Papers)",
    month = jul,
    year = "2025",
    address = "Vienna, Austria",
    publisher = "Association for Computational Linguistics",
    url = "https://aclanthology.org/2025.acl-long.123/",
}

@inproceedings{nguyen2020zero,
  title={The Zero Resource Speech Benchmark 2021: Metrics and baselines for unsupervised spoken language modeling},
  author={Nguyen, Tu Anh and de Seyssel, Maureen and Roz{\'e}, Patricia and Rivi{\`e}re, Morgane and Kharitonov, Evgeny and Baevski, Alexei and Dunbar, Ewan and Dupoux, Emmanuel},
  booktitle={Neur{IPS} Workshop on Self-Supervised Learning for Speech and Audio Processing},
  year={2020}
}

@inproceedings{maimon2025salmon,
  title={Salmon: A Suite for Acoustic Language Model Evaluation},
  author={Maimon, Gallil and Roth, Amit and Adi, Yossi},
  booktitle={ICASSP 2025 - IEEE International Conference on Acoustics, Speech and Signal Processing (ICASSP)},
  year={2025},
  pages={1--5},
  doi={10.1109/ICASSP49660.2025.10888561}
}

@article{warstadt-etal-2020-blimp-benchmark,
    title = "{BL}i{MP}: The Benchmark of Linguistic Minimal Pairs for {E}nglish",
    author = "Warstadt, Alex  and
      Parrish, Alicia  and
      Liu, Haokun  and
      Mohananey, Anhad  and
      Peng, Wei  and
      Wang, Sheng-Fu  and
      Bowman, Samuel R.",
    editor = "Johnson, Mark  and
      Roark, Brian  and
      Nenkova, Ani",
    journal = "Transactions of the Association for Computational Linguistics",
    volume = "8",
    year = "2020",
    address = "Cambridge, MA",
    publisher = "MIT Press",
    url = "https://aclanthology.org/2020.tacl-1.25/",
    doi = "10.1162/tacl_a_00321",
    pages = "377--392",
}

@inproceedings{le-godais-etal-2017-comparing,
    title = "Comparing Character-level Neural Language Models Using a Lexical Decision Task",
    author = {Le Godais, Ga{\"e}l  and
      Linzen, Tal  and
      Dupoux, Emmanuel},
    editor = "Lapata, Mirella  and
      Blunsom, Phil  and
      Koller, Alexander",
    booktitle = "Proceedings of the 15th Conference of the {E}uropean Chapter of the Association for Computational Linguistics: Volume 2, Short Papers",
    month = apr,
    year = "2017",
    address = "Valencia, Spain",
    publisher = "Association for Computational Linguistics",
    url = "https://aclanthology.org/E17-2020/",
}

@inproceedings{zellers-etal-2019-hellaswag,
    title = "{H}ella{S}wag: Can a Machine Really Finish Your Sentence?",
    author = "Zellers, Rowan  and
      Holtzman, Ari  and
      Bisk, Yonatan  and
      Farhadi, Ali  and
      Choi, Yejin",
    editor = "Korhonen, Anna  and
      Traum, David  and
      M{\`a}rquez, Llu{\'i}s",
    booktitle = "Proceedings of the 57th Annual Meeting of the Association for Computational Linguistics",
    month = jul,
    year = "2019",
    address = "Florence, Italy",
    publisher = "Association for Computational Linguistics",
    url = "https://aclanthology.org/P19-1472/",
    doi = "10.18653/v1/P19-1472",
    pages = "4791--4800",
}

@inproceedings{panayotov2015librispeech,
  title={Librispeech: an ASR corpus based on public domain audio books},
  author={Panayotov, Vassil and Chen, Guoguo and Povey, Daniel and Khudanpur, Sanjeev},
  booktitle={Acoustics, Speech and Signal Processing (ICASSP), 2015 IEEE International Conference on},
  pages={5206--5210},
  year={2015},
  organization={IEEE}
}

@misc{marin8bretro,
  title={Marin 8B Retrospective},
  author={{Marin Community}},
  year={2025},
  howpublished={\url{https://marin.readthedocs.io/en/latest/reports/marin-8b-retro/}},
  note={Accessed: 2025-10-30}
}

@article{hibiki,
  title={High-Fidelity Simultaneous Speech-To-Speech Translation},
  author={Tom Labiausse and Laurent Mazar{\'e} and Edouard Grave and Alexandre D{\'e}fossez and Neil Zeghidour},
  booktitle={Forty-second International Conference on Machine Learning},
  year={2025},
  url={https://openreview.net/forum?id=fgjN8B6xVX}
}

@inproceedings{unity,
    title = "{U}nit{Y}: Two-pass Direct Speech-to-speech Translation with Discrete Units",
    author = "Inaguma, Hirofumi  and
      Popuri, Sravya  and
      Kulikov, Ilia  and
      Chen, Peng-Jen  and
      Wang, Changhan  and
      Chung, Yu-An  and
      Tang, Yun  and
      Lee, Ann  and
      Watanabe, Shinji  and
      Pino, Juan",
    editor = "Rogers, Anna  and
      Boyd-Graber, Jordan  and
      Okazaki, Naoaki",
    booktitle = "Proceedings of the 61st Annual Meeting of the Association for Computational Linguistics (Volume 1: Long Papers)",
    month = jul,
    year = "2023",
    address = "Toronto, Canada",
    publisher = "Association for Computational Linguistics",
    url = "https://aclanthology.org/2023.acl-long.872/",
    doi = "10.18653/v1/2023.acl-long.872",
    pages = "15655--15680",
}

@misc{datadecide,
      title={DataDecide: How to Predict Best Pretraining Data with Small Experiments}, 
      author={Ian Magnusson and Nguyen Tai and Ben Bogin and David Heineman and Jena D. Hwang and Luca Soldaini and Akshita Bhagia and Jiacheng Liu and Dirk Groeneveld and Oyvind Tafjord and Noah A. Smith and Pang Wei Koh and Jesse Dodge},
      year={2025},
      eprint={2504.11393},
      archivePrefix={arXiv},
      primaryClass={cs.LG},
      url={https://arxiv.org/abs/2504.11393}, 
}

@article{zheng2025rosettaspeech,
  title={RosettaSpeech: Zero-Shot Speech-to-Speech Translation from Monolingual Data},
  author={Zheng, Zhisheng and Sun, Xiaohang and Dinh, Tuan and Yanamandra, Abhishek and Jain, Abhinav and Liu, Zhu and Hadap, Sunil and Bhat, Vimal and Aggarwal, Manoj and Medioni, Gerard and others},
  journal={arXiv preprint arXiv:2511.20974},
  year={2025}
}
\bibliographystyle{icml2026}

\newpage
\appendix
\onecolumn

\section{Tokenizer and Data Formatting}
\label{sec:appendix_method}

This section provides details on the tokenizer and data formatting. For model architecture and audio codec details, see \S\ref{sec:experimental_setup}.

\subsection{Tokenizer}
\label{sec:appendix_tokenizer}

SODA uses a custom tokenizer built on top of the Marin tokenizer\footnote{\url{https://huggingface.co/stanford-crfm/marin-tokenizer}}, which is based on Llama 3's tokenizer, consisting of:
\begin{itemize}
    \item \textbf{Text tokens}: 128,256 tokens from the base Marin tokenizer
    \item \textbf{Audio tokens}: 16,384 tokens ($8$ codebooks $\times$ $2{,}048$ codebook size) for Mimi codec
    \item \textbf{Special tokens}: 4 addition tokens, \texttt{<|text\_start|>}, \texttt{<|text\_end|>}, \texttt{<|audio\_start|>}, \texttt{<|audio\_end|>}
    \item \textbf{Total vocabulary}: 144,644 tokens
\end{itemize}

For audio tokens, we use direct codebook mapping with no BPE merges. Initially, we experimented with a merge list of up to 128K audio token merges but observed only $\sim$10\% token reduction. Given low reduction, we use 0 merges for simplicity.

\textit{Note}: For warm-start experiments (\S\ref{sec:warm_cold_comparison}) that initialize from Qwen3 weights, we use the Qwen3 tokenizer instead of Marin to match the pre-trained embeddings.

\subsection{Data Formatting and Interleaving}
\label{sec:appendix_formatting}

A key design choice in SODA is utterance-level interleaving of text and audio tokens (illustrated in Figure~\ref{fig:token_types}). Each audio file (e.g., a full YouTube video) is processed through the dataset pipeline (Yodas/Emilia/MLS), which applies VAD to segment audio into chunks. For a document with $N$ chunks (i.e., speech segments), we construct two training variants:

\textbf{Text-first format}:
\begin{verbatim}
<|begin_of_text|>
<|text_start|>text_1<|text_end|>
<|audio_start|>audio_1<|audio_end|>
<|text_start|>text_2<|text_end|>
<|audio_start|>audio_2<|audio_end|>
...
<|text_start|>text_N<|text_end|>
<|audio_start|>audio_N<|audio_end|>
<|end_of_text|>
\end{verbatim}

\textbf{Audio-first format}:
\begin{verbatim}
<|begin_of_text|>
<|audio_start|>audio_1<|audio_end|>
<|text_start|>text_1<|text_end|>
<|audio_start|>audio_2<|audio_end|>
<|text_start|>text_2<|text_end|>
...
<|audio_start|>audio_N<|audio_end|>
<|text_start|>text_N<|text_end|>
<|end_of_text|>
\end{verbatim}

Here, \texttt{chunk\_i} and \texttt{chunk\_(i+1)} are adjacent audio segments from the same document. Adjacent chunks may be temporally contiguous or separated by silence (depending on VAD); we do not treat them differently. Although this process is supposed to yield speech-only data, we observe many audio segments containing non-speech content (e.g., background noise, music, etc.), which allows our model to learn broader audio knowledge although not evaluated in this work.

\textbf{Sequence packing}: We pack sequences to the maximum context length (4,096 tokens for most experiments) with no padding, maximizing training efficiency. Multiple documents may be packed into a single training sequence, separated by \texttt{<|end\_of\_text|>} tokens. Also, a single document longer than the maximum context length may be split into multiple training sequences.

This dual-format strategy ensures the model learns bidirectional cross-modal capabilities: audio$\rightarrow$text (ASR) and text$\rightarrow$audio (TTS), as well as audio continuation and text continuation, all within a single pre-training stage.

\subsection{Training Hyperparameters}
\label{sec:appendix_hyperparameters}

We provide training hyperparameters for reproducibility. Our small-scale (in \S~\ref{sec:design_choice}) and large-scale (in \S\ref{sec:largescale}) experiments use the same base configuration and optimizer settings, scaled appropriately for model size.

\textbf{Optimizer.} We use AdamW with $\beta_1 = 0.98$, $\beta_2 = 0.98$, $\epsilon = 10^{-16}$, weight decay $0.033$, and gradient clipping with max norm $1.0$. We apply a small $z$-loss with weight $10^{-4}$ to the output logits, following our previous work that improved pretraining stability text LLMs. Note that the Chameleon work \cite{chameleon} also showed that $z$-loss is particularly effective for multi-modal models on training stability.

\textbf{Learning Rate Schedule.} We use the Warmup-Stable-Decay (WSD) schedule~\cite{hu2024minicpm} with 10\% warmup, 70\% stable, and 20\% linear decay phases. The base learning rate is $0.003$ at batch size 256 and model width 1024. For other configurations, we scale the learning rate as:
\begin{equation}
    \text{lr} = 0.003 \times \sqrt{\frac{\text{batch\_size}}{256}} \times \sqrt{\frac{1024}{\text{width}}}
\end{equation}
where width is the model's hidden dimension (e.g., 512 for 135M, 1024 for 600M, 2048 for 1.7B, 2560 for 4B).

\textbf{Batch Size and Sequence Length.} For large-scale experiments (in \S\ref{sec:largescale}), we use a sequence length of 4096 tokens and batch size of 512 for most experiments. The number of training steps is computed as $\lceil \text{tokens} / (\text{batch\_size} \times \text{seq\_len}) \rceil$.

\textbf{Hardware.} All experiments in this paper are conducted on Google Cloud TPU v5p. For small-scale runs, e.g., design choice ablations in \S\ref{sec:design_choice} and most of IsoFLOP runs in \S\ref{sec:scaling}, we use v5p-8 or v5p-16 depending on model size. For large-scale training (\S\ref{sec:largescale}), we use v5p-32 (135M), v5p-64 (600M), v5p-128 (1.7B), and v5p-256 (4B).

For our IsoFLOP sweep runs of 64 models, please refer to exact configurations and hyperparameters outlined in our code at \texttt{experiments/audio/isoflop\_audio\_sweep.py}.

\section{Training Data}
\label{sec:appendix_data_statistics}

We provide detailed statistics and descriptions for all datasets used in this work. Table~\ref{tab:data_statistics} summarizes the key properties.

\begin{table}[h]
    \centering
    \caption{Training data statistics. Hours and tokens are for English subsets used in our experiments. Token counts assume 100 tokens/sec for audio (8 Mimi codebooks at 12.5 Hz) and standard BPE tokenization for text. *Estimated from disk size. $\dagger$Token counts include both audio-first and text-first formats (i.e., unrepeated token counts would be half of this value).}
    \label{tab:data_statistics}
    \small
    \begin{tabular}{@{}lllccl@{}}
    \toprule
    \textbf{Dataset} & \textbf{Split} & \textbf{Type} & \textbf{Hours} & \textbf{$\sim$Tokens} & \textbf{Source} \\
    \midrule
    Yodas~\cite{li2023yodas} &English & Speech & 164K & $\sim$131B$^\dagger$ & YouTube (100+ langs) \\
    Emilia~\cite{he2024emilia} &English+Yodas-English & Speech & 139K & $\sim$110B$^\dagger$ & YouTube (6 langs) \\
    MLS~\cite{mls} &English & Speech & 44.5K & $\sim$35B$^\dagger$ & LibriVox audiobooks \\
    \midrule
    Nemotron-CC~\cite{su-etal-2025-nemotron} &HQ-Actual & Text & -- & $\sim$220B* & Web (CC filtered) \\
    \bottomrule
    \end{tabular}
\end{table}

\textbf{Speech Data.} We select the largest publicly available speech corpora with utterance-level transcriptions:

\textit{Yodas}~\cite{li2023yodas} is one of the largest open speech datasets, containing over 370K hours across 100+ languages sourced from YouTube with automatic transcripts. For our main experiments, we use the English subset ($\sim$164K hours, $\sim$131B tokens). We note that in our preliminary experiment (SODA-prelim), we use 8 languages: English (131B tokens), Spanish (35B), French (17B), German (11B), Thai (0.5B), Hindi (0.4B), Arabic (0.3B), and Chinese (0.2B), totaling $\sim$196B tokens. Later we wanted to isolate the impact of multilingual training, so we only used English data for SODA-prelim. The dataset offers diverse speaking styles, recording conditions, and utterance lengths, making it well-suited for training general-purpose audio models.

\textit{Emilia}~\cite{he2024emilia} is a large-scale multilingual dataset ($\sim$101K hours across 6 languages) specifically designed for speech generation. It emphasizes spontaneous, natural speech with diverse speaking styles. We use the English subset ($\sim$140K hours). Emilia's focus on spontaneity complements Yodas's scale.

\textit{MLS} (Multilingual LibriSpeech)~\cite{mls} is derived from LibriVox audiobooks, containing $\sim$45K hours of English read speech. While MLS is a high-quality curated corpus, we find in \S\ref{sec:speech_data_selection} that it performs poorly for our setting due to: (i) uncased, unpunctuated transcripts creating distribution mismatch with standard text, and (ii) fixed-length 10--20 second chunks lacking utterance length diversity. We therefore exclude MLS from our final training recipe.

\textbf{Text Data.} To improve semantic understanding and world knowledge, we incorporate high-quality text-only data:

\textit{Nemotron-CC}~\cite{su-etal-2025-nemotron} is a large-scale filtered web corpus derived from Common Crawl, widely used in LLM pre-training. We use the high-quality actual split, containing around 220B tokens (approximated from its disk size).

\textbf{Final Data Mixture.} Based on our design choice investigations in \S\ref{sec:design_choice}, we use the following sampling weights for the IsoFLOP analysis (\S\ref{sec:scaling}) and large-scale SODA training (\S\ref{sec:largescale}): 5\% text (Nemotron) and 95\% speech (Yodas + Emilia). Within the speech portion, we sample proportionally to dataset size (shown in Table~\ref{tab:data_statistics}), resulting in 51.6\% of Yodas English, 28.8\% of Emilia-YODAS English, and 14.6\% of Emilia English. Since pre-training uses standard instance sampling and packing, these ratios closely reflect the actual token distribution during training.

\section{Evaluation Benchmarks}
\label{sec:appendix_evaluation}

We evaluate SODA models across four categories: speech semantic knowledge, speech acoustic knowledge, text knowledge, and cross-modal skills. This section describes each benchmark and evaluation protocol.

\subsection{Speech Semantic Knowledge}

\textbf{sBLIMP} \cite{nguyen2020zero} evaluates grammatical knowledge from speech utterances. The benchmark contains minimal pairs of spoken sentences that differ in grammaticality (e.g., ``The cat sleeps'' vs. ``The cat sleep''). Models assign likelihood to each sentence, and accuracy is computed as the percentage of pairs where the grammatical sentence receives higher likelihood. We use the full sBLIMP test set and report zero-shot accuracy. Random baseline: 50\%.

\textbf{sWUGGY} \cite{nguyen2020zero} tests phonotactic knowledge---the ability to distinguish real words from phonologically plausible non-words in speech. Each item contains a real word and a pseudoword (e.g., ``oscillation'' vs. ``odenacia''). Models assign likelihood to each audio token, and accuracy is the percentage where the real word receives higher likelihood. We use the full sWUGGY test set. Random baseline: 50\%.

\subsection{Speech Acoustic Knowledge}
\textbf{Salmon} \cite{maimon2025salmon} evaluates acoustic modeling quality through acoustic consistency tasks. For each test sample, the benchmark compares the model's likelihood of an original recording against a modified version where an acoustic property (e.g., speaker identity, background noise, or room acoustics) changes mid-utterance. The score is the percentage of samples where the model correctly assigns higher likelihood to the consistent recording. The metric ranges from 0--100\%, with higher values indicating better acoustic modeling. Random baseline: 50\%.

\subsection{Text Knowledge}

\textbf{tBLIMP} \cite{warstadt-etal-2020-blimp-benchmark} is the original text-only version of BLIMP, evaluating grammatical knowledge from written text. The evaluation protocol is identical to sBLIMP but uses text tokens instead of audio tokens. Random baseline: 50\%.

\textbf{tWUGGY} \cite{le-godais-etal-2017-comparing} is the original text-only version of WUGGY, testing lexical knowledge from written text. The evaluation protocol is identical to sWUGGY but uses text tokens. Random baseline: 50\%.

\textbf{HellaSwag} \cite{zellers-etal-2019-hellaswag} evaluates commonsense reasoning and world knowledge. Each item presents a context and four possible continuations; models must select the most plausible one. We use the validation set and compute zero-shot accuracy by selecting the continuation with the highest likelihood. Random baseline: 25\%.

\subsection{Cross-Modal Skills}

\textbf{ASR (Speech-to-Text)} measures speech-to-text transcription quality. We evaluate the ASR ability on LibriSpeech (test-clean) \cite{panayotov2015librispeech}, reporting Word Error Rate (WER).

\textbf{TTS (Text-to-Speech)} evaluates zero-shot speech generation quality using the seed-tts-eval benchmark based on its official repo in \url{https://github.com/BytedanceSpeech/seed-tts-eval}. We measure two metrics: (1) \textbf{WER} (intelligibility): generated speech is transcribed with \textit{Whisper-large-v3} and compared against the input text; (2) \textbf{SIM} (speaker similarity): cosine similarity between \textit{WavLM-large} speaker embeddings of generated speech and reference audio. We use the English split of Seed-TTS-Eval.

\subsection{Validation Loss Metrics}

Beyond task-specific benchmarks, in Section~\ref{sec:validation_loss_metric}, we compute validation negative log-likelihood (NLL) on the following held-out data. All NLL metrics are computed using the standard cross-entropy loss with the model's vocabulary. Lower NLL indicates better predictive performance.

\textbf{LibriSpeech} (audio tokens + transcript tokens): We compute NLL on LibriSpeech test-clean, measuring the model's predictive performance on this English speech corpus. We report NLL on all tokens (semantic + acoustic + text) from audio-first formatted data. Note that Section~\ref{sec:validation_loss_metric} also investigates other options for different NLL metrics.

\textbf{Paloma-c4en} (text tokens): We compute NLL on the C4-en subset of Paloma, a standardized text-only perplexity benchmark. Following the Marin-8B work \cite{marin8bretro}, we use Paloma-c4en as our primary validation metric for tracking text modeling progress, as it has been shown to be a reliable indicator of general text understanding capabilities. We report NLL on text tokens only.

\section{Design Choice Investigations: Additional Details}
\label{sec:appendix_design_choice}

This section provides additional details for the design choice investigations in \S\ref{sec:design_choice}.

\subsection{SODA-prelim: Preliminary Experiment}
\label{sec:appendix_prelim}

Before conducting our design choice investigations, we first validated whether training a decoder-only transformer from scratch on discrete audio tokens yields meaningful capabilities. It was unclear whether semantic and acoustic tokens from Mimi, interleaved at the utterance level with text tokens, could produce coherent speech generation or cross-modal understanding without pre-trained encoders.

\textbf{Setup.} We trained a 600M parameter model on 500B tokens from Yodas2, covering eight languages (English, Spanish, French, German, Thai, Hindi, Arabic, and Chinese). This model, which we call SODA-prelim, uses a Warmup-Stable-Decay learning rate schedule, enabling us to study the impact of different data sources during the annealing (fast cooling-down) phase without full retraining~\cite{blakeney2024spark}.

\textbf{Findings.} The results (shown in Table~\ref{tab:main_table} in the main text) reveal both strengths and limitations:
\begin{itemize}
    \item \textit{Acoustic understanding}: SODA-prelim achieves 69.4\% on Salmon, outperforming SpiritLM-base (57.2\%) and SpiritLM-Expr (67.1\%), confirming that joint semantic-acoustic modeling is effective.
    \item \textit{Cross-modal skills}: SODA-prelim demonstrates functional ASR (22.0\% WER zero-shot, improving to 15.2\% with 2-shot prompting) and TTS (9.2\% WER), validating the unified semantic, acoustic, and text token training.
    \item \textit{Semantic understanding}: Performance is poor (sBLIMP: 50.9\%, sWUGGY: 57.8\%), suggesting the model struggles to acquire linguistic structure from audio alone.
    \item \textit{Text knowledge}: Near random performance (HellaSwag: 26.2\%), indicating that the initial recipe fails to transfer general knowledge effectively.
\end{itemize}

These findings motivated the subsequent design choice investigations in \S\ref{sec:design_choice}: speech data selection (\S\ref{sec:speech_data_selection}), text mixture inclusion (\S\ref{sec:text_data_mixture_ratio}), and token composition ablations (\S\ref{sec:token_composition}). SODA-prelim also serves as the base checkpoint for the annealing experiments used in speech data selection.

\subsection{Experimental Setup Rationale}
\label{sec:appendix_design_choice_setup}

The three subsections in \S\ref{sec:design_choice} use different experimental setups, each tailored to the specific question being addressed:

\textbf{\S\ref{sec:speech_data_selection} (Speech Data Selection)} uses \textit{annealing experiments}~\cite{blakeney2024spark}, where we branch from a pre-trained checkpoint (SODA-prelim) during the learning rate decay phase to compare data sources. We expect this to be the most reliable approach for comparing data quality, as it isolates the effect of data while keeping all other factors constant. However, it requires a pre-trained checkpoint and is expensive per comparison. We also validate our findings with small-scale runs (150M models, 10B tokens); see \S\ref{sec:appendix_speech_data} for full results.

\textbf{\S\ref{sec:text_data_mixture_ratio} (Text Ratio)} requires sweeping many ratios (0\%--50\%) to find the optimal balance. Given that small-scale runs (150M, 10B tokens) yielded similar findings to annealing in our previous experiment in \S\ref{sec:speech_data_selection}, we adopt this cheaper setup here, allowing efficient exploration of the ratio space.

\textbf{\S\ref{sec:token_composition} (Token Composition)} ablates token types (semantic-only vs.\ semantic+acoustic vs.\ semantic+acoustic+text). Annealing is not applicable here because different token types require different model configurations from the start. With only three configurations to compare, we can afford a larger compute budget per run. We use 1.7B models trained on 30B tokens ($3 \times 10^{20}$ FLOPs), a configuration informed by our concurrent IsoFLOP experiments (\S\ref{sec:scaling}).

\subsection{Speech Data Selection: Full Results}
\label{sec:appendix_speech_data}

Table~\ref{tab:speech_data_selection} presents the full results of our speech data selection experiments from \S\ref{sec:speech_data_selection}. We evaluate three speech corpora---MLS, Emilia, and Yodas---using two experimental setups: (1) annealing experiments branching from SODA-prelim, and (2) small-scale runs training 150M models on 10B tokens from scratch.

\begin{table}[h]
    \centering
    \caption{Speech data selection results. Top: SODA-prelim baseline (full training). Middle: annealing experiments comparing different data sources. Bottom: small-scale validation runs (150M parameters, 10B tokens). MLS performs poorly on cross-modal tasks despite being a curated audiobook corpus, while Yodas and Emilia show complementary strengths. ASR = LibriSpeech-clean-test, TTS=seed-tts-eval.}
    \label{tab:speech_data_selection}
    \small
    \setlength{\tabcolsep}{3pt}
    \begin{tabular}{@{}lccccccccc@{}}
    \toprule
    & \multicolumn{2}{c}{\textbf{Semantic}} & \textbf{Acoustic} & \multicolumn{2}{c}{\textbf{Text}} & \textbf{ASR} & \multicolumn{2}{c}{\textbf{TTS}} \\
    \cmidrule(lr){2-3} \cmidrule(lr){4-4} \cmidrule(lr){5-6} \cmidrule(lr){7-7} \cmidrule(lr){8-9}
    \textbf{Data} & sWUGGY$\uparrow$ & sBLIMP$\uparrow$ & Salmon$\uparrow$ & tWUGGY$\uparrow$ & tBLIMP$\uparrow$ & WER$\downarrow$ & WER$\downarrow$ & SIM$\uparrow$ \\
    \midrule
    \multicolumn{9}{l}{\textit{SODA-prelim}} \\
    Yodas (multilingual) & 57.8 & 50.9 & 69.4 & 71.3 & 69.0 & 22.0 & 9.2 & 0.516 \\
    \midrule
    \multicolumn{9}{l}{\textit{Annealing from SODA-prelim (at final 20\% cooldown steps)}} \\
    MLS   (English) & 56.8 & 50.6 & 70.3 & 69.1 & 72.2 & 92.6 & 35.7 & 0.366 \\
    Emilia (English) & 58.5 & 51.0 & 69.8 & 58.5 & 70.5 & 7.8  & 6.1  & 0.557 \\
    Yodas (English)  & 57.8 & 51.0 & 69.8 & 73.8 & 68.6 & 18.8 & 12.4 & 0.502 \\
    \midrule
    \multicolumn{9}{l}{\textit{Small-scale runs (150M, 10B tokens)}} \\
    MLS (English)   & 54.9 & 49.7 & 69.5 & 57.2 & 65.6 & 105.6 & -- & -- \\
    Emilia (English) & 55.8 & 49.4 & 67.9 & 48.6 & 67.1 & 26.8  & -- & -- \\
    Yodas (English) & 54.9 & 49.6 & 69.4 & 64.9 & 60.9 & 54.1  & -- & -- \\
    \bottomrule
    \end{tabular}
\end{table}

Both experimental setups yield consistent conclusions: MLS performs poorly on cross-modal tasks (ASR WER: 92.6\% in annealing, 105.6\% in small-scale), while Emilia and Yodas show viable performance. The small-scale runs, despite being noisier due to limited compute, support our finding from annealing experiments that MLS should be omitted from the training mixture. We attribute MLS's poor performance to (i) uncased, unpunctuated transcripts creating distribution mismatch with standard text, and (ii) fixed-length 10--20 second chunks lacking utterance length diversity. Between Emilia and Yodas, Emilia excels at TTS (WER: 6.1\%, SIM: 0.557), while Yodas provides better text knowledge (tWUGGY: 73.8\%). We therefore combine both in our final recipe.

\subsection{Text-Only Data Inclusion: Full Results}
\label{sec:appendix_nemotron}

\begin{figure}[h]
    \centering
    \includegraphics[width=0.99\textwidth]{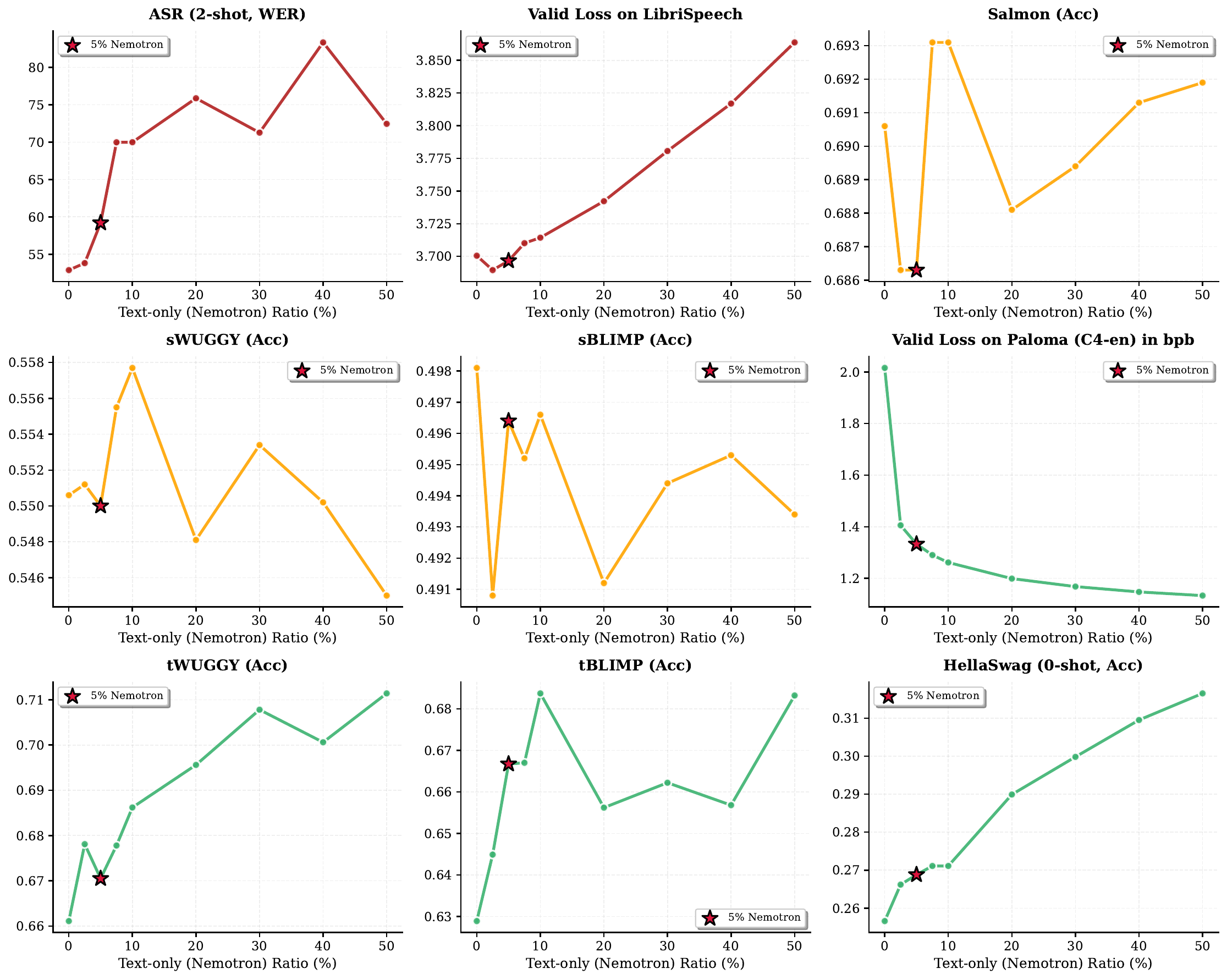}
    \caption{Full results of text-injection ratio sweep (0\%--90\% Nemotron text data). Colors indicate the trend as text percentage increases: \textcolor{red}{\textbf{Red}} (Cross-Modal skills: ASR, TTS) shows degradation in performance as text ratio increases beyond 5\%. \textcolor{orange}{\textbf{Orange}} (Semantic and Acoustic understanding: sWUGGY, sBLIMP, Salmon) indicates little variation across text ratios, suggesting these capabilities are robust to text inclusion. \textcolor{green}{\textbf{Green}} (Text Knowledge tasks: tWUGGY, tBLIMP, HellaSwag) shows monotonic improvement as text percentage increases, with notable gains from 0\% to 2.5\%.}
    \label{fig:nemotron_sweep_appendix}
\end{figure}

Figure~\ref{fig:nemotron_sweep_appendix} shows the full results of the text-injection ratio sweep described in \S~\ref{sec:text_data_mixture_ratio}. The 3$\times$3 grid displays all evaluated metrics across three categories.

\newpage
\text{}
\newpage
\section{NLL Correlation with Downstream Metrics: Full Results}
\label{sec:appendix_nll}

\begin{figure}[h!]
    \centering
    \includegraphics[width=0.99\textwidth]{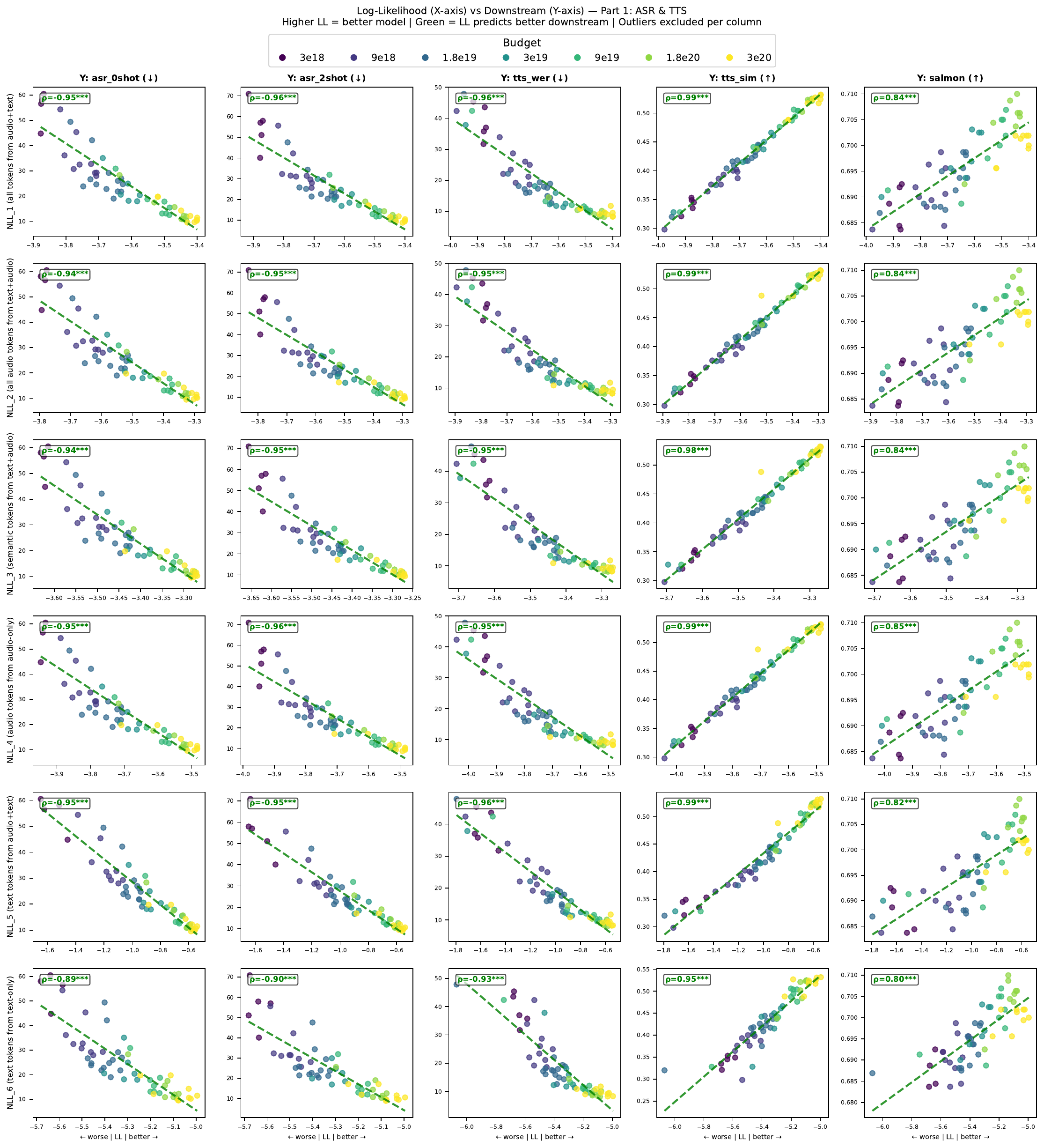}
    \caption{Validation NLL (audio-first) versus downstream task performance (Part 1). Each subplot shows one metric across 64 models from the IsoFLOP study, with Spearman $\rho$ reported. Cross-modal metrics (ASR, TTS) show strong correlation with NLL.}
    \label{fig:nll_correlation_full1}
\end{figure}

\begin{figure}[h!]
    \centering
    \includegraphics[width=0.99\textwidth]{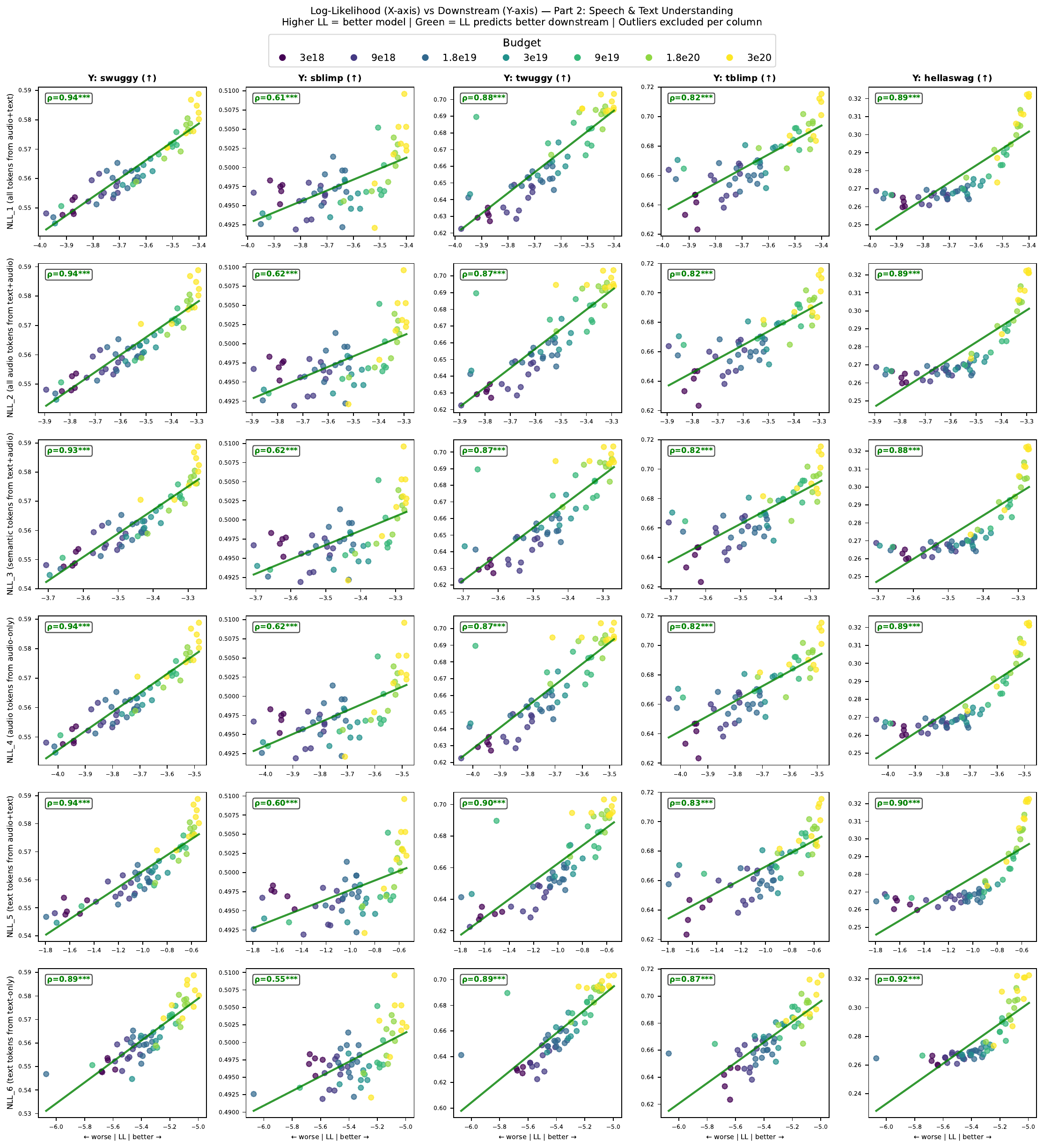}
    \caption{Validation NLL (audio-first) versus downstream task performance (Part 2). Speech semantic and acoustic metrics show continued improvement without saturation. Text knowledge tasks (tWUGGY, HellaSwag) show emergence patterns with accelerated improvement at lower NLL values.}
    \label{fig:nll_correlation_full2}
\end{figure}

Figures~\ref{fig:nll_correlation_full1} and~\ref{fig:nll_correlation_full2} show the full correlation analysis between validation NLL and all downstream metrics across 64 models from our IsoFLOP study (\S~\ref{sec:validation_loss_metric}). Each subplot shows a scatter plot with Spearman rank correlation ($\rho$) and a fitted trend line.

\textbf{Choice of NLL Metric.} Our interleaved training format (semantic, acoustic, and text tokens in audio-first or text-first order) admits multiple ways to compute validation NLL. In Figures~\ref{fig:nll_correlation_full1} and~\ref{fig:nll_correlation_full2}, different rows correspond to the following NLL metrics: (1) all tokens from audio+text data, (2) all audio tokens from text+audio data, (3) semantic tokens from text+audio data, (4) audio tokens from audio-only data, (5) text tokens from audio+text data, (6) text tokens from text-only data.

Table~\ref{tab:nll_spearman} quantifies the Spearman correlations (absolute values) between each NLL metric and downstream tasks. While all variants show similar correlations, we select \textbf{NLL on all tokens from audio+text data (Option 1)} as our primary metric because it provides the best balance: it achieves near-best correlation with speech tasks (Avg\_Speech = 0.895, only 0.001 behind the best) while maintaining strong correlation with text tasks (Avg\_Text = 0.864). In contrast, Option 4 (audio-only NLL) is marginally better for speech (0.896) but worse for text (0.859), while Option 6 (text-only NLL) is best for text (0.894) but substantially worse for speech (0.848). For a general-purpose model targeting both modalities, Option 1 offers the most balanced proxy.

\begin{table}[h]
    \centering
    \caption{Spearman correlation (absolute values) between NLL metrics and downstream tasks across 64 IsoFLOP models. Avg\_Speech averages 7 speech metrics; Avg\_Text averages 3 text metrics. Option 1 achieves the best balance between speech and text correlations.}
    \label{tab:nll_spearman}
    \small
    \begin{tabular}{@{}lcccccccccccc@{}}
    \toprule
    \textbf{NLL} & \textbf{ASR-0} & \textbf{ASR-2} & \textbf{TTS-W} & \textbf{TTS-S} & \textbf{Sal.} & \textbf{sWUG} & \textbf{sBLI} & \textbf{tWUG} & \textbf{tBLI} & \textbf{Hella} & \textbf{Avg-Sp} & \textbf{Avg-Tx} \\
    \midrule
    NLL\_1 & 0.959 & 0.967 & 0.955 & 0.989 & 0.841 & 0.941 & 0.614 & 0.879 & 0.824 & 0.887 & 0.895 & 0.864 \\
    NLL\_2 & 0.956 & 0.962 & 0.953 & 0.986 & 0.842 & 0.936 & 0.621 & 0.874 & 0.821 & 0.887 & 0.894 & 0.861 \\
    NLL\_3 & 0.954 & 0.959 & 0.952 & 0.985 & 0.842 & 0.929 & 0.620 & 0.866 & 0.819 & 0.884 & 0.892 & 0.856 \\
    NLL\_4 & 0.959 & 0.965 & 0.952 & 0.987 & 0.845 & 0.937 & 0.623 & 0.871 & 0.819 & 0.886 & 0.896 & 0.859 \\
    NLL\_5 & 0.962 & 0.961 & 0.960 & 0.986 & 0.820 & 0.944 & 0.602 & 0.899 & 0.827 & 0.904 & 0.891 & 0.877 \\
    NLL\_6 & 0.903 & 0.908 & 0.933 & 0.949 & 0.802 & 0.892 & 0.552 & 0.891 & 0.873 & 0.917 & 0.848 & 0.894 \\
    \bottomrule
    \end{tabular}
\end{table}

\newpage
\text{}
\newpage
\text{}
\newpage
\section{Validation Loss Prediction for Over-Trained Models}
\label{sec:appendix_parametric}

This section analyzes whether our scaling laws can predict the validation loss of the final SODA models, which are heavily over-trained relative to the compute-optimal frontier.

\begin{table}[h]
    \centering
    \caption{Training regime for SODA models. $D^*$ is the compute-optimal token count predicted by our scaling laws (\S\ref{sec:isoflop_analysis}). Over-training factor indicates how many times more tokens we train on compared to optimal.}
    \label{tab:training_regime}
    \small
    \begin{tabular}{@{}ccc@{}}
    \toprule
    \textbf{Model Size} & \textbf{$D^*$} & \textbf{Over-training Factor} \\
    \midrule
    135M & 0.53B  & 940$\times$ \\
    600M & 5.55B  & 90$\times$ ($\sim$ Llama3's) \\
    1.7B & 28.6B  & 18$\times$ ($\sim$ Llama2's) \\
    4B   & 110B   & 4.5$\times$ \\
    \bottomrule
    \end{tabular}
\end{table}

Table~\ref{tab:training_regime} summarizes the over-training regime for SODA models. Smaller models are heavily over-trained (e.g., 135M at 940$\times$ $D^*$, comparable to aggressive over-training in recent LLM practice), while the 4B model approaches the compute-optimal frontier.

\subsection{Compute-Optimal Frontier Extrapolation}

\begin{figure}[h]
    \centering
    \includegraphics[width=0.7\textwidth]{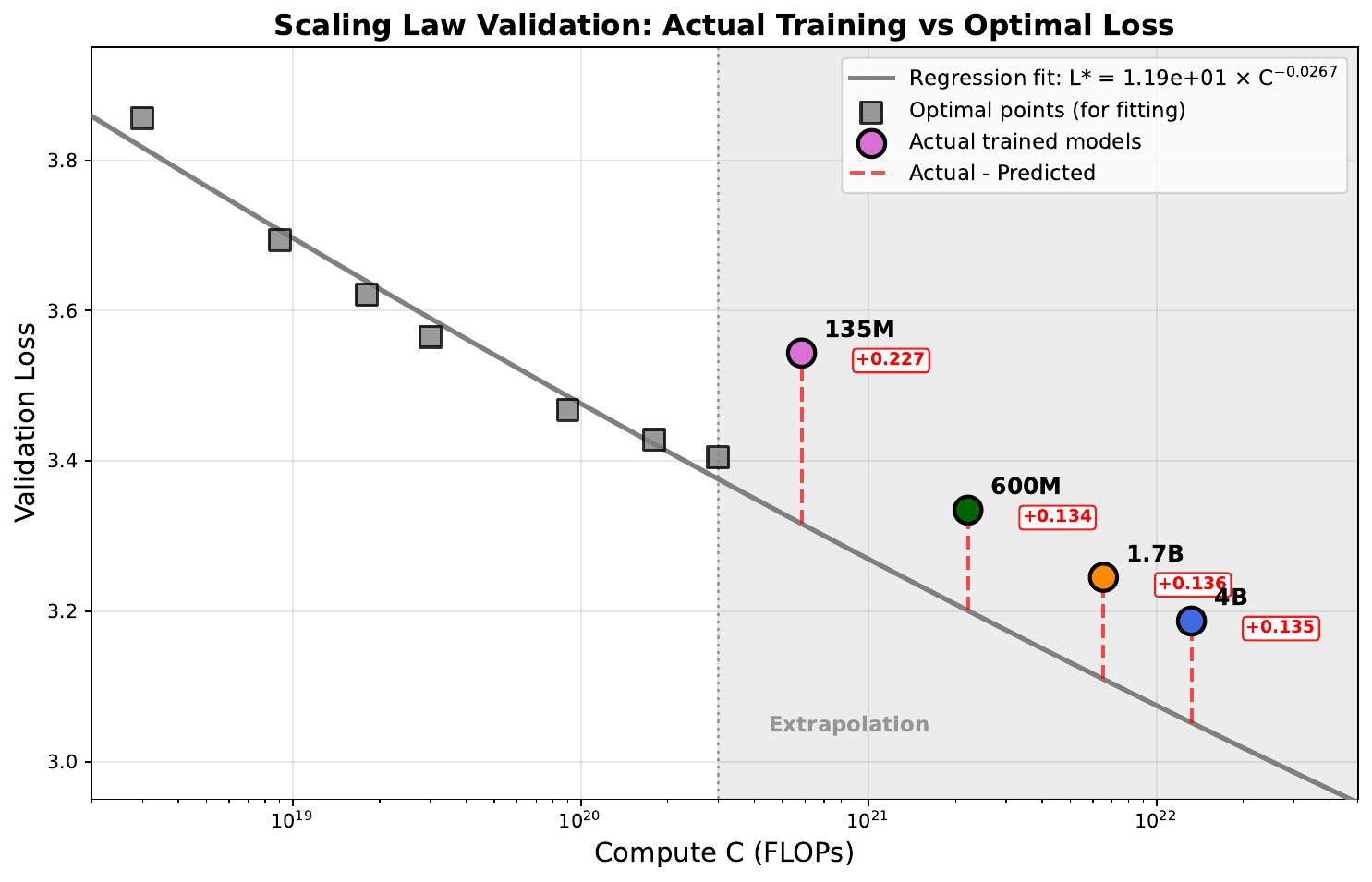}
    \caption{Validation loss vs. compute for SODA models. Grey squares show compute-optimal models from \S\ref{sec:scaling}. We fit a power-law regression $L^* = a \cdot C^b$ on these seven points and extrapolate to higher compute (grey line). Colored circles show final SODA runs.}
    \label{fig:scaling_results}
\end{figure}

Figure~\ref{fig:scaling_results} compares over-trained SODA models against the extrapolated compute-optimal frontier. We fit a power-law regression $L^* = a \cdot C^b$ on the seven compute-optimal points from our IsoFLOP study and extrapolate to higher compute budgets. All SODA models achieve higher loss than the optimal extrapolated prediction at equivalent compute, with smaller models deviating more due to heavier over-training (e.g., 135M at 940$\times$ $D^*$). This is expected: over-trained models use smaller-than-optimal model sizes for their compute budget, trading off loss for inference efficiency.

\subsection{Parametric Loss Prediction}

An alternative approach is to fit a parametric equation that captures loss as a function of both model size $N$ and training tokens $D$, allowing us to directly predict loss for over-trained configurations. We fit the parametric equation from Chinchilla~\cite{chinchilla}:
\begin{equation}
    L(N, D) = E + \frac{A}{N^\alpha} + \frac{B}{D^\beta}
\end{equation}
where $N$ is the model size (parameters), $D$ is the number of training tokens, and $E$ represents the irreducible loss. We fit this equation on all 64 data points from our IsoFLOP sweep (not just the compute-optimal points), obtaining:
\begin{equation}
    L = 3.169 + \frac{215886}{N^{0.684}} + \frac{4750}{D^{0.439}} \quad (R^2 = 0.983)
\end{equation}

\textbf{Derived Scaling Exponents}. The parametric fit also provides an alternative derivation of scaling exponents. From the Chinchilla derivation, the compute-optimal model size and data scale as $N^* \propto C^{\beta/(\alpha+\beta)}$ and $D^* \propto C^{\alpha/(\alpha+\beta)}$. Using our fitted $\alpha = 0.684$ and $\beta = 0.439$, we obtain:
\begin{align}
    N^* &\propto C^{0.391} \quad \text{(vs. } C^{0.367} \text{ in IsoFLOP fit in Eqn~\ref{eq:scaling_law_N} in \S\ref{sec:isoflop_analysis})} \\
    D^* &\propto C^{0.609} \quad \text{(vs. } C^{0.579} \text{ in IsoFLOP fit in Eqn~\ref{eq:scaling_law_D} in \S\ref{sec:isoflop_analysis})}
\end{align}
These derived exponents are reasonably close to the empirically fitted values from \S\ref{sec:isoflop_analysis}, with both approaches showing $D^*/N^* \approx 1.57$. This asymmetry (data scaling faster than model size) is consistent with the lower information density of discrete audio tokens compared to text.

\begin{figure}[h]
    \centering
    \includegraphics[width=0.7\textwidth]{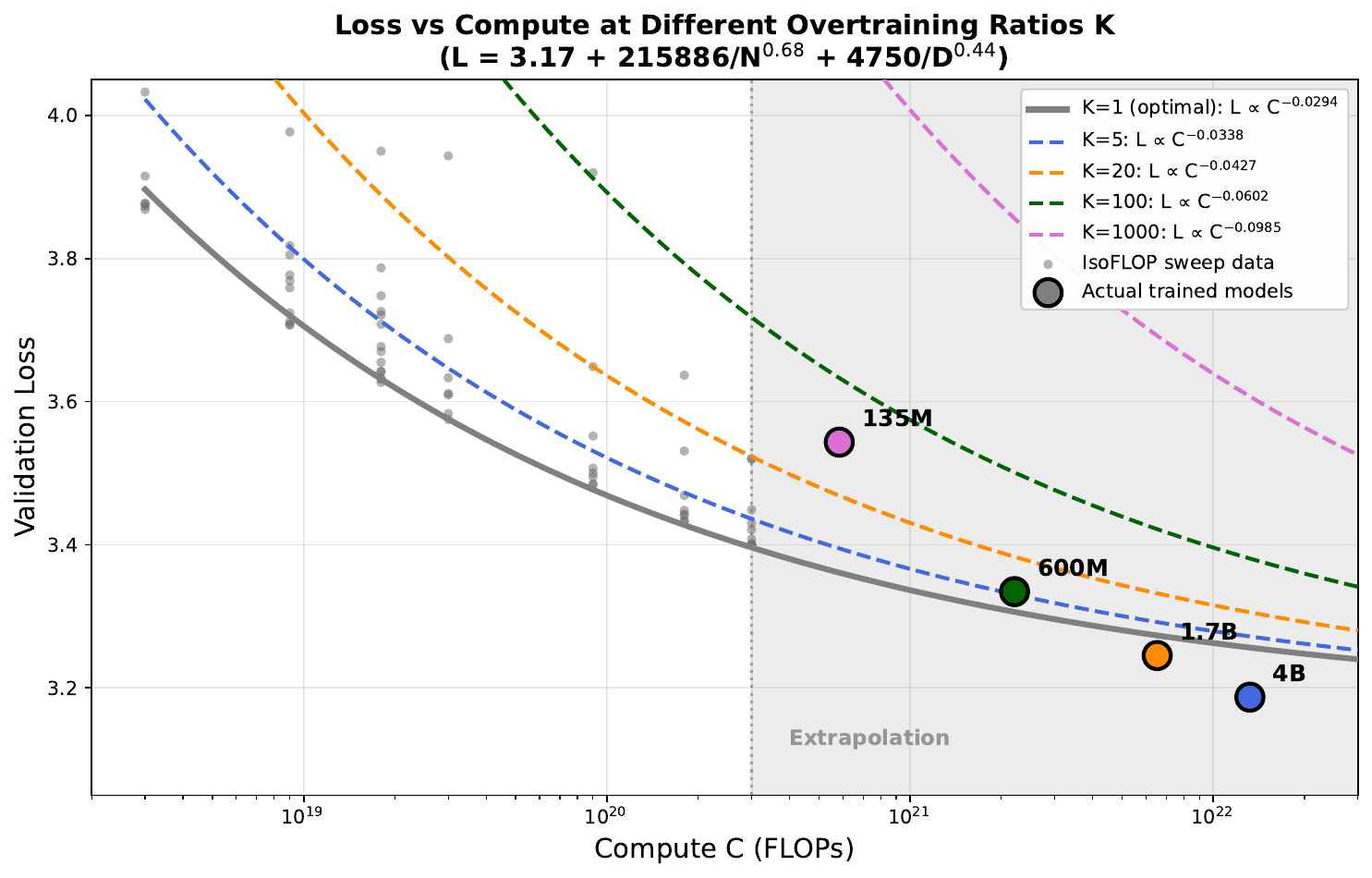}
    \caption{Loss vs. compute at different overtraining ratios $K$ using the fitted parametric equation. Grey points show IsoFLOP sweep data. Colored curves show predicted loss at $K=1$ (optimal), $K=5$, $K=20$, and $K=100$. Colored circles show actual SODA models (135M, 600M, 1.7B, 4B), which achieve lower losses than predicted. The discrepancy shows that SODA models outperform the parametric prediction, suggesting favorable scaling dynamics or lower irreducible loss at larger scales.}
    \label{fig:parametric_fit}
\end{figure}

\textbf{Loss Predictions}. Figure~\ref{fig:parametric_fit} shows predictions at different overtraining ratios $K = D/D^*$, where $K=1$ is compute-optimal and higher $K$ indicates more overtraining (smaller model trained on more data). The actual SODA models (135M--4B) achieve losses \textit{lower} than predicted, even lower than the optimal extrapolated prediction for the 1.7B and 4B models. This systematic under-prediction when extrapolating from the IsoFLOP regime ($3 \times 10^{18}$--$3 \times 10^{20}$ FLOPs) to higher compute ($6 \times 10^{20}$--$1.3 \times 10^{22}$ FLOPs) could be due to scaling law extrapolations diverging in the far tails of the compute distribution. The offset suggests that either scaling exponents evolve favorably at larger scales, or the irreducible loss $E$ is lower than estimated from the smaller-scale regime.

\subsection{Summary}

Neither the compute-optimal extrapolation nor the parametric fit very accurately predicts validation loss for our final SODA runs. The compute-optimal approach predicts lower losses than observed (expected, since over-trained models are suboptimal), while the parametric approach predicts higher losses than observed. This discrepancy suggests that scaling behavior may evolve favorably beyond the IsoFLOP regime, or that our smaller-scale fits do not fully capture dynamics at larger compute budgets. Despite these prediction challenges, we note that downstream task performance scales reliably with NLL (see Figure~\ref{fig:nll_correlation} and \S\ref{sec:largescale}), making NLL a useful proxy even when absolute loss predictions are imperfect.

\section{Warm-Start vs. Cold-Start: Training Trajectories}
\label{sec:appendix_warmstart}

\begin{figure}[h!]
    \centering
    \begin{subfigure}[b]{0.48\textwidth}
        \centering
        \includegraphics[width=\textwidth]{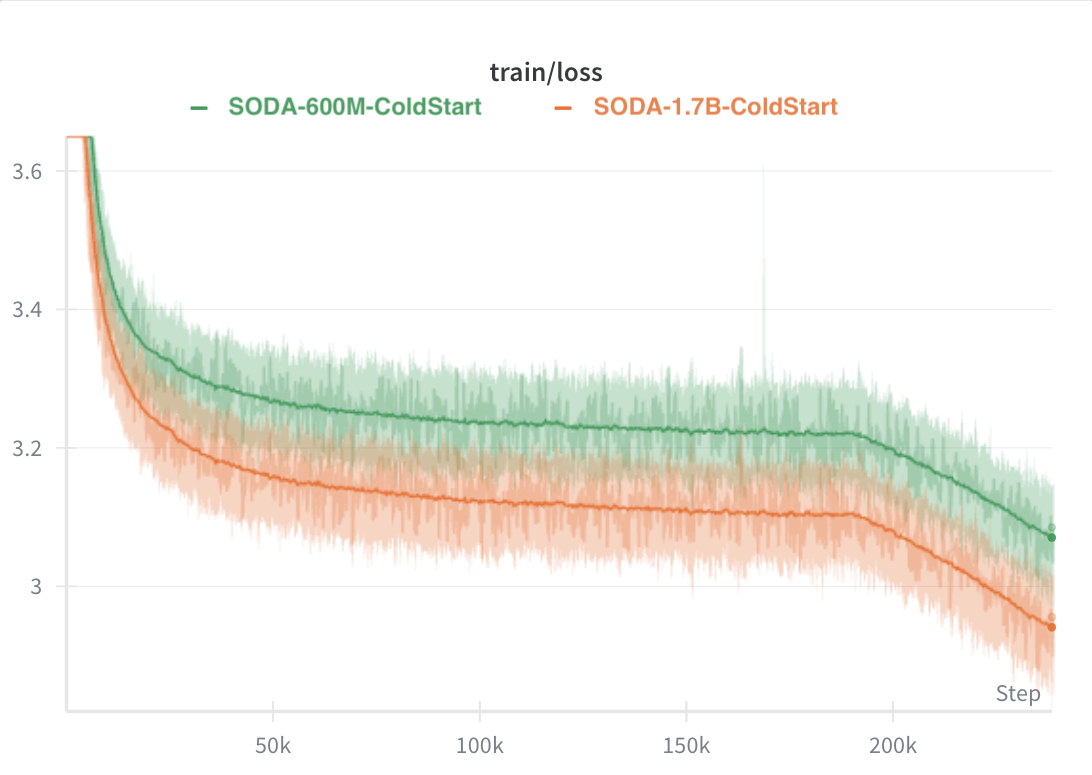}
        \caption{Cold-Start: smooth improvement.}
    \end{subfigure}
    \hfill
    \begin{subfigure}[b]{0.48\textwidth}
        \centering
        \includegraphics[width=\textwidth]{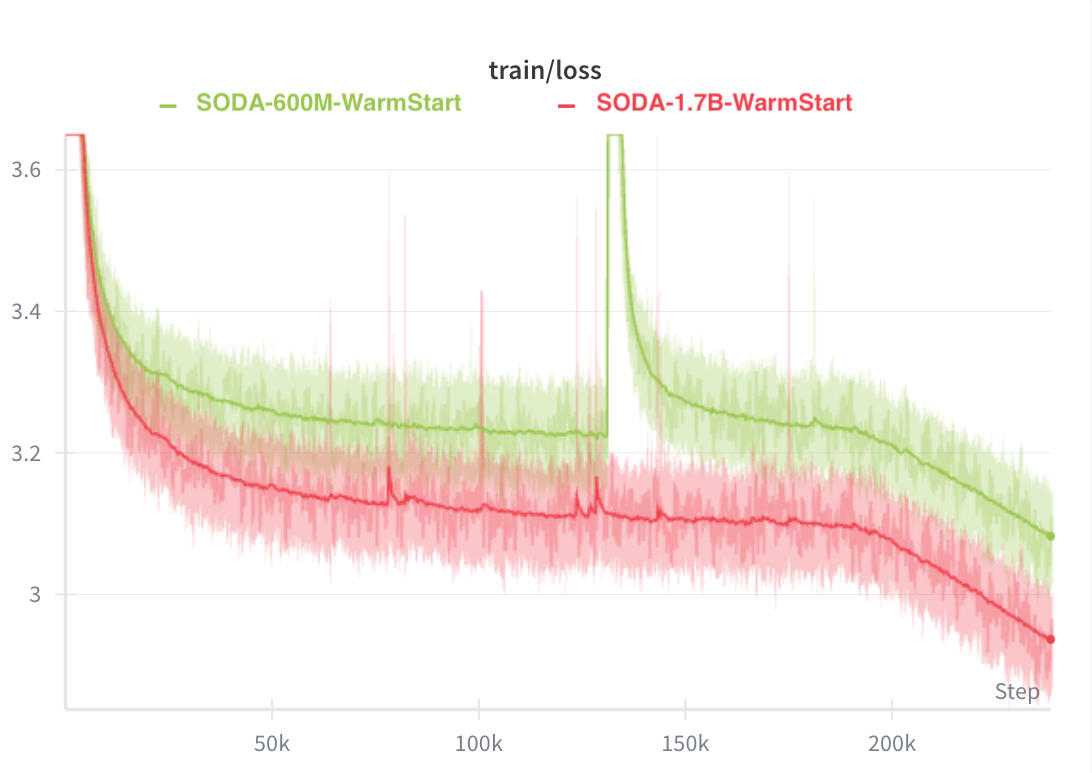}
        \caption{Warm-Start: frequent spikes; 600M has one large spike.}
    \end{subfigure}
    \caption{\textbf{Training loss curve} (extracted from W\&B) comparison between Cold-Start and Warm-Start at 600M and 1.7B scales. Warm-Start exhibits instability with many small spikes throughout training for both model sizes. The 600M Warm-Start run shows a particularly large spike around 135K steps. We note that with our limited compute budget, we are able to train only one run for each model configuration. However, we note that better regularization or different hyperparameters could stabilize the training process of Warm-Start.}
    \label{fig:train_loss_stability}
\end{figure}

\begin{figure}[h!]
    \centering
    \includegraphics[width=0.99\textwidth]{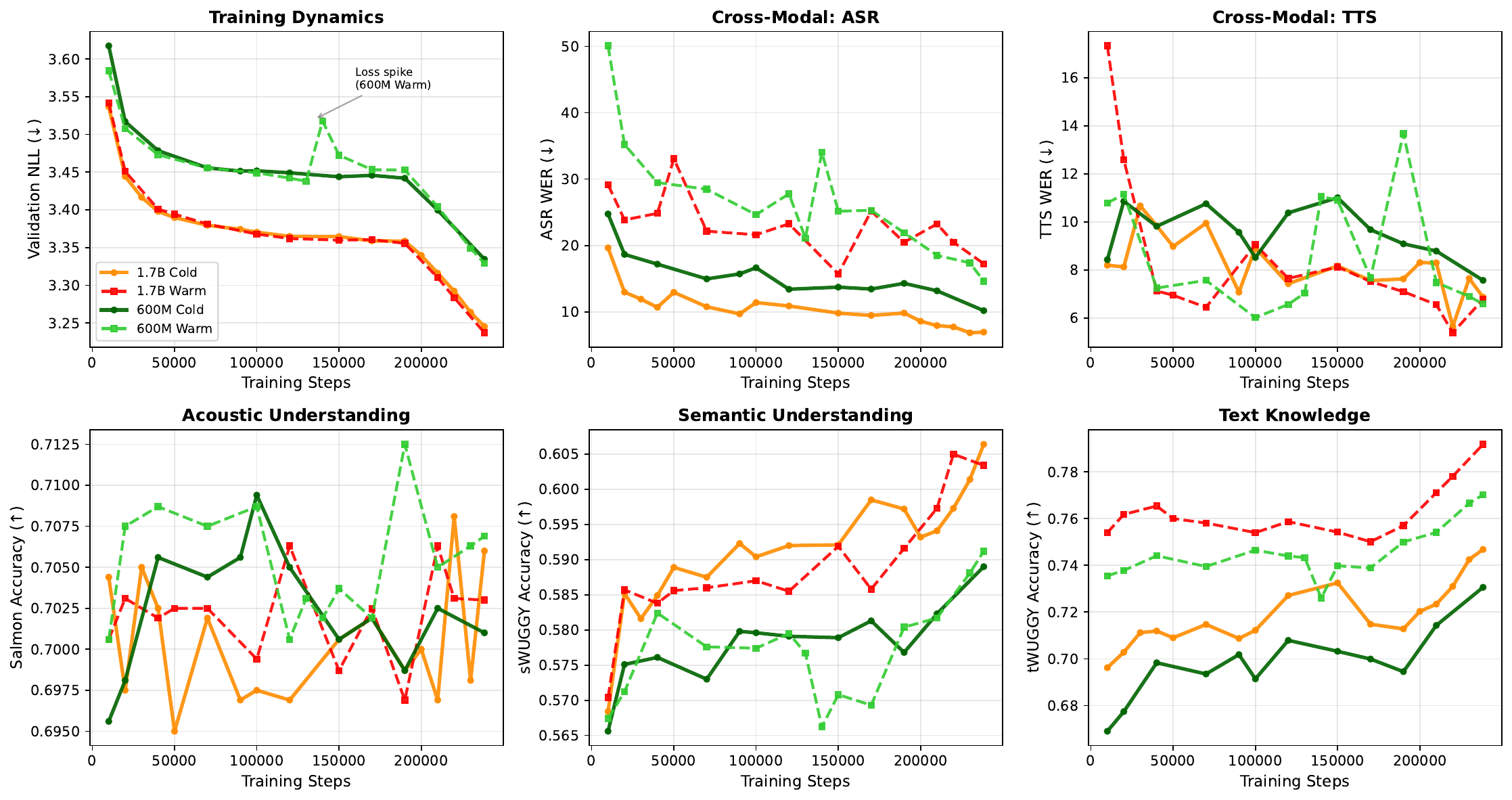}
    \caption{Warm-Start vs. Cold-Start training comparison at 600M and 1.7B scales. \textbf{Top row}: Training dynamics (NLL) and cross-modal skills (ASR, TTS). Cold-Start achieves lower ASR-WER throughout training, while TTS converges similarly. The NLL plot highlights a loss spike at 135K steps for the 600M Warm-Start run. \textbf{Bottom row}: Understanding tasks. Salmon (acoustic) and sWUGGY (semantic) show similar trajectories regardless of initialization, suggesting these are learned fresh. tWUGGY (text knowledge) shows Warm-Start maintaining a persistent advantage throughout training, with Cold-Start never catching up.}
    \label{fig:warm_cold_comparison}
\end{figure}

Table~\ref{tab:warm_cold_comparison} presents the final evaluation results, and Figure~\ref{fig:warm_cold_comparison} shows the full training trajectories comparing Warm-Start (initialized from Qwen3) versus Cold-Start (trained from scratch) at 600M and 1.7B scales. Each subplot tracks a different metric across 500B tokens of training. Figure~\ref{fig:train_loss_stability} compares training loss stability between the two initialization strategies.

\begin{table*}[h]
    \caption{Warm-Start vs. Cold-Start training comparison at 600M and 1.7B scales. Final evaluation results are shown.}
    \label{tab:warm_cold_comparison}
    \centering
    \small
    \setlength{\tabcolsep}{1.8pt}
    \renewcommand{\arraystretch}{0.975}
    \begin{tabular}{@{}lcccccccccc@{}}
    \toprule
    & \multicolumn{2}{c}{\textbf{Speech (Semantic)}} & \textbf{Speech (Acoustic)} & \multicolumn{3}{c}{\textbf{Text (Knowledge)}} & \multicolumn{3}{c}{\textbf{Cross-Modal}} \\
    \cmidrule(lr){2-3} \cmidrule(lr){4-4} \cmidrule(lr){5-7} \cmidrule(lr){8-10}
    \textbf{Model} & \textbf{sBLIMP}$\uparrow$ & \textbf{sWUGGY}$\uparrow$ & \textbf{Salmon}$\uparrow$ & \textbf{tBLIMP}$\uparrow$ & \textbf{tWUGGY}$\uparrow$ & \textbf{HellaS}$\uparrow$ & \textbf{ASR}$\downarrow$ & \textbf{TTS$_{\text{WER}}$}$\downarrow$ & \textbf{TTS$_{\text{SIM}}$}$\uparrow$ \\
    \midrule
    SODA-600M (cold-start) & 51.2 & 58.9 & 70.1 & 70.7 & 73.1 & 35.8 & 10.2 & 7.6 & 0.555 \\
    SODA-600M (warm-start) & 51.1 & 59.1 & 70.7 & 70.8 & 77.0 & 36.3 & 14.6 & 6.6 & 0.559 \\
    \midrule
    SODA-1.7B (cold-start) & 51.4 & 60.6 & 70.6 & 70.3 & 74.7 & 44.5 & 7.0  & 6.9 & 0.560 \\
    SODA-1.7B (warm-start) & 51.8 & 60.3 & 70.3 & 71.0 & 79.2 & 47.1 & 17.3 & 6.8 & 0.557 \\

    \bottomrule
    \end{tabular}
\end{table*}

\section{Speech-to-Speech Translation: Dataset and Fine-tuning Details}
\label{sec:appendix_s2st}

This section provides additional details for the S2ST experiment in \S\ref{sec:s2st}.

\textbf{Dataset: CVSS-T}. We use CVSS-T~\cite{cvss}, a multilingual speech-to-speech translation corpus derived from CoVoST 2. CVSS-T contains $\sim$1.9K hours of paired speech across 21 source languages translating to English. Crucially, the target English speech is synthesized to \textit{preserve the source speaker's voice} using a voice-cloning TTS system, making this a voice-preserving S2ST benchmark. The training set contains $\sim$557K examples.

\textbf{Evaluation}. Due to the computational cost of evaluating the full CVSS-T test set ($\sim$84K examples), we subsample 800 examples: 200 each from Spanish (es), French (fr), German (de), and 200 from the remaining 18 languages (``other''). We report: (1) \textbf{ASR-BLEU}: BLEU score computed on ASR-transcribed model outputs against reference translations; (2) \textbf{SIM}: Speaker similarity between the generated and source speech. For both metrics, we use the same evaluation setup as seed-tts-eval (as used in our TTS evaluation), with Whisper-large-v3 for ASR and WavLM-large for speaker similarity.

\textbf{Comparison Context}. Direct numerical comparison with prior work is difficult due to different evaluation splits. However, for context: Unity~\cite{unity} reports Es$\rightarrow$En BLEU of 18.2 (S2TT+TTS cascade), Translatotron 2~\cite{translatotron2} achieves 25.4 (task-specific architecture). On voice preservation, Hibiki~\cite{hibiki} achieves SIM $\approx 0.41$ and Seamless~ achieves SIM $\approx 0.30$. Our SODA models achieve SIM $\approx 0.47$, competitive with or exceeding these task-specific systems.

\textbf{Fine-tuning Hyperparameters}. We fine-tune on the full CVSS-T training set (557K examples) for 5 epochs using:
\begin{itemize}[leftmargin=*]
    \setlength\itemsep{-0.2em}
    \item Learning rate: cosine schedule with peak $2 \times 10^{-5}$
    \item Batch size: 64
    \item Sequence length: 4096 tokens
    \item The same optimizer settings as pre-training (Appendix~\ref{sec:appendix_hyperparameters})
\end{itemize}

\textbf{Data Formatting}. We format S2ST as interleaved next-token prediction: [source audio tokens] $\rightarrow$ [source text] $\rightarrow$ [target text] $\rightarrow$ [target audio tokens]. This is the same format used for ASR and TTS, with the addition of source$\rightarrow$target text translation in between.


\end{document}